\documentclass[twocolumn]{aastex631}

\usepackage{amsmath}
\usepackage{amssymb}
\usepackage{xcolor}

\usepackage[whole]{bxcjkjatype}

\newcommand{\revise}[1]{\textcolor{black}{#1}}

\newcommand{\Fearth}{F_{\rm XUV, \oplus}}
\newcommand{\FXUV}{F_{\rm XUV}}
\newcommand{\Texo}{T_{\rm exo}}

\shorttitle{Survival of terrestrial N$_2$-O$_2$ atmosphere in violent XUV environments}
\shortauthors{Nakayama et al.}
\graphicspath{{./}{figures/}}

\begin{document}

\title{
Survival of Terrestrial N$_2$-O$_2$ Atmospheres in Violent XUV 
Environments \revise{through} Efficient Atomic Line Radiative Cooling 
}

\correspondingauthor{Akifumi Nakayama}
\email{anakayama@rikkyo.ac.jp}

\author[0000-0002-0998-0434]{Akifumi Nakayama}
\affiliation{Department of Physics, College of Science, Rikkyo University, 3-34-1 Nishi-Ikebukuro, Toshima-ku, Tokyo 171-8501, Japan}
\affiliation{Department of Earth and Planetary Science, Graduate School of Science, The University of Tokyo, 7-3-1 Hongo, Bunkyo-ku, Tokyo 113-0033, Japan}

\author[0000-0002-5658-5971]{Masahiro Ikoma}
\affiliation{Division of Science, National Astronomical Observatory of Japan (NAOJ), 2-21-1 Osawa, Mitaka, Tokyo 181-8558, Japan}
\affiliation{Department of Astronomical Science, The Graduate University for Advanced Studies (SOKENDAI), 2-21-1 Osawa, Mitaka, Tokyo 181-8558, Japan}

\author[0000-0001-5685-9736]{Naoki Terada}
\affiliation{Department of Geophysics, Graduate School of Science, Tohoku University, 6-3 Aramaki-aza-Aoba, Aoba-ku, Sendai, Miyagi 980-8578, Japan}



\begin{abstract}
Atmospheres play a crucial role in planetary habitability.
Around M dwarfs and young Sun-like stars, planets receiving the same insolation as the present-day Earth are exposed to intense stellar X-rays and extreme-ultraviolet (XUV) radiation. 
This study explores the fundamental question of whether the atmosphere of present-day Earth could survive in such harsh XUV environments.
Previous theoretical studies suggest that stellar XUV irradiation is sufficiently intense to remove such atmospheres completely on short timescales. 
In this study, we develop a new upper-atmospheric model and re-examine the thermal and hydrodynamic responses of the thermospheric structure of an Earth-like N$_2$-O$_2$ atmosphere, on an Earth-mass planet, to an increase in the XUV irradiation.
Our model includes the effects of radiative cooling via electronic transitions of atoms and ions, known as atomic line cooling, in addition to the processes accounted for by previous models.
We demonstrate that atomic line cooling dominates over the hydrodynamic effect at XUV irradiation levels greater than several times the present level of the Earth. 
Consequentially, the atmosphere's structure is kept almost hydrostatic, and its escape remains sluggish even at XUV irradiation levels up to a thousand times that of the Earth at present. 
Our estimates for the Jeans escape rates of N$_2$-O$_2$ atmospheres suggest that these 1-bar atmospheres survive in early active phases of Sun-like stars.
Even around active late M dwarfs, N$_2$-O$_2$ atmospheres could escape significant thermal loss on timescales of gigayears.
These results give new insights into the habitability of terrestrial exoplanets and the Earth's climate history.
\end{abstract}

\keywords{Earth, planets and satellites: terrestrial planets, planets and satellites: atmospheres}


\section{Introduction} \label{sec:intro}


Progress in exoplanet science has led to a growing interest in the astronomical and planetary science communities as to whether Earth-like habitable planets commonly exist beyond the solar system. 
Recent exoplanet surveys have already detected about two dozen likely rocky planets similar in insolation to the Earth, including TRAPPIST-1~d, e, and f \citep[][]{Gillon2016,Gillon2017}. 
Such ``temperate'' exoplanets identified thus far are orbiting late-type dwarfs, or M dwarfs, which have a lower effective temperature and luminosity than the Sun. 
Around M dwarfs, X-ray and extreme-ultraviolet, hereafter collectively referred to as XUV, account for a larger proportion of the stellar bolometric luminosity compared to Sun-like stars \citep[e.g.,][]{Scalo2007, West2008}. 
This means that planets with the same insolation receive more intense XUV around M dwarfs than around Sun-like stars.
A fundamental question is then raised as to whether the present-day Earth could remain a habitable planet in such harsh environments. 
Of particular interest may be the retention of planetary atmospheres, since the atmospheres of rocky planets play a crucial role in their climates and habitability.

While several physical processes drive atmospheric escape \citep[e.g.,][for a recent review]{Tian2015, Gronoff2020}, 
thermal escape driven by stellar XUV irradiation is capable of significantly eroding the atmosphere. 
Such atmospheric loss is thought to have occurred enormously in the early history of the Earth \citep[]{Sekiya1980, Watson1981}, 
because stellar XUV luminosity diminishes by orders of magnitude over 
the course of stellar evolution, through stellar spin-down and a decline in coronal activity \citep[e.g.,][]{Ribas2005, Tu2015}.
M dwarfs are known to evolve slowly, maintaining much stronger XUV emission than the Sun on geological timescales \citep[e.g.,][]{Scalo2007, West2008}. 
This suggests that planets currently located in the habitable zone \citep[see][for the traditional definition]{Kasting1993} around \revise{M dwarfs} are subject to continual atmospheric erosion.
\revise{Climate simulations show that the location and width of the habitable zone depend on the heterogeneity of surface conditions such as water distributions \citep[e.g.,][]{Abe2011, Leconte2013}; the actual location of the habitable zone is, however, not important in this study, since we focus on atmospheric escape.}

Previous theoretical studies \citep[][]{Kulikov2007,Tian2008a,Tian2008b} concluded that the terrestrial atmosphere is vulnerable to high-level XUV irradiation estimated for the early Earth and the temperate planets orbiting \revise{M dwarfs}. 
Specifically, \citet[]{Tian2008a, Tian2008b} demonstrated that an atmosphere with the same elemental abundances as the Earth, hereafter referred to as an Earth-like N$_2$-O$_2$ atmosphere, lapses into a hydrodynamic state once the XUV irradiation exceeds approximately five times the present level experienced by Earth. 
For high-level XUV irradiation, the thermospheric temperature is as high as $\sim$~10,000~K. An extremely hot thermosphere is attributed to the inefficiency of infrared (IR) radiative cooling via molecular vibrational-rotational transitions.
Recent hydrodynamic simulations for an Earth-like N$_2$-O$_2$ atmosphere irradiated by XUV from the very active young Sun at 4.4~Ga yield extremely high rates of atmospheric escape so that the 1-bar atmosphere is completely lost within 0.1~Myr \citep[]{Johnstone2019}. 
These simulations also showed that the mass loss occurs in an energy-limited fashion \citep[for the definition, see][]{Sekiya1980,Watson1981} with a heating efficiency of almost unity.
This is because the hot escaping upper atmosphere contributes little to the efficiency of the molecular IR cooling occurring below the homopause.

However, these previous theoretical models disregard the cooling effects of
the line emission that occurs through the transition of electronic states in atoms and ions, known as atomic line cooling, which are excited via collisions with ambient thermal electrons and neutral and ion species, although considering the effects of superthermal electrons (or photoelectrons) generated via photoionization and/or direct precipitation \citep[e.g.,][]{Tian2008b, Garcia-sage2017}. 
Given that atmospheric models without atomic line cooling reproduce the observed structure of the Earth’s atmosphere \citep[e.g.,][]{Johnstone2018}, atomic line cooling contibutes little to Earth’s thermosphere at present.
However atomic line cooling is known to be a common effective process in several hot astronomical fields, such as hydrogen Lyman-$\alpha$ cooling in hot Jupiter atmospheres \citep[e.g.,][]{Murray2009}, 
hydrogen Ly$\alpha$ and H$\alpha$ cooling in accreting flows for young gas giants \citep[][]{Aoyama2019}, and line emission of hydrogen and metals at shock fronts in molecular clouds \citep[e.g.,][]{Hollenbach1989}. 
Recently \citet[]{Ito2021} 
investigated the hydrodynamic escape of the rocky-vapor atmosphere on top of the magma ocean of close-in super-Earths and showed that atomic line cooling from the metals, Na, Mg, Si, and O, has a great impact on the upper atmospheric structure. 
Consequently they found that the rocky-vapor atmosphere is much more resistant to stellar XUV radiation than previously thought. 
This finding motivates us to incorporate the effect of atomic line cooling in the thermal structure and stability of an Earth-like N$_2$-O$_2$ atmosphere.

To date, we have had a poor understanding of how effective such atomic line cooling is in high-XUV environments for an Earth-like N$_2$-O$_2$ atmosphere, corresponding to 
terrestrial exoplanets in the habitable zone around \revise{M dwarfs} and the early Earth.
Since metals such as O, N, and their ions have smaller energy intervals between the electronic states (or finer structure intervals) than hydrogen, atomic excitation occurs at relatively low energies or temperatures.
Thus, atomic line cooling is expected to be effective in lowering the thermospheric temperature and thereby suppressing thermal escape for strongly XUV-irradiated Earth-like atmospheres. 
In this study we aim to confirm and quantify this inference by developing new thermospheric models of an Earth-like N$_2$-O$_2$ atmosphere with a focus on
the energy balance including atomic line cooling. 

The remainder of this paper is organized as follow: 
In Section~\ref{sec:model}, we describe the upper-atmospheric model for terrestrial planets. 
In Section~\ref{sec:result}, we show the results of the response of the atmospheric structure to an increase in XUV irradiation with a focus on the importance of atomic line cooling. 
In Section~\ref{sec:discussion}, we compare our results with those from previous studies and discuss the stability of N$_2$-O$_2$ atmospheres of Earth-mass planets in the habitable zone around Sun-like stars and \revise{M dwarfs}.
Finally, we conclude this paper in Section~\ref{sec:conclusion}.

\section{Model and Calculation Method} \label{sec:model}
We develop a radial 1D structural model of the upper atmosphere of a terrestrial planet. 
In this study, we assume an Earth-mass planet with an atmosphere that contains the four elements, H, C, N, and O, with relative abundances the same as those in the present atmosphere of the Earth.
In Section~\ref{sec:basic equation}, we present the basic equations that determine the atmospheric structure. 
In Section~\ref{sec:chemical reaction}, we summarize the thermo- and photochemical reactions, which include 47 neutral and ion species composed from the four elements. 
In Section~\ref{sec:diffusion conduction}, we describe diffusion and conduction. 
In Section~\ref{sec:cooling heating}, we describe our new treatment of radiative cooling and heating, followed by a brief description of hydrodynamic cooling in Section~\ref{sec:advection}. 
In Sections~\ref{sec:numerical} and \ref{sec:parameters}, we summarize our numerical procedure and parameters, including stellar XUV spectra, respectively. 
A benchmark test is done in Appendix~\ref{Ap:benchmark}.

\subsection{Basic equations} \label{sec:basic equation}
We integrate the time-dependent equation of continuity,
\begin{align} 
    \frac{\partial n_j}{\partial t} + \frac{1}{r^2} \frac{\partial}{\partial r} \left[  r^2 n_j(u + u_j) \right] &= S_j  \label{eq:continuous}, 
\end{align}
where $t$ is the time, $r$ is the radial distance measured from the planet's center, 
$u$ is the bulk velocity, and 
$n_j$, $u_j$, and $S_j$ are 
the number density, diffusive velocity, and source/sink term (including thermochemical and photochemical reactions) of species $j$, respectively. The energy equation is given by
\begin{align} 
    \frac{\partial e}{\partial t} = Q^{\rm chem} + Q^{\rm uv} + Q^{\rm pe} + Q^{\rm diff} + Q^{\rm rad} - Q^{\rm hy}, \label{eq:energy}
\end{align}
where $e$ is the energy density,
$Q^{\rm chem}$ is the heating rate through chemical reactions,
$Q^{\rm uv}$ is that of photodissociation,
$Q^{\rm pe}$ is that of collision with photoelectrons, 
$Q^{\rm diff}$ is that of thermal diffusion,
$Q^{\rm rad}$ is the total rate of molecular radiation and absorption, and atomic line cooling,
and 
$Q^{\rm hy}$ is the rate of hydrodynamic 
cooling. 
Note that no kinetic energy of the bulk flow is included because a steady-state motion of the atmospheric gas is assumed in this study.

We consider a steady-state motion of an inviscid, ideal gas. 
The equation of motion (or the Euler equation) is written in terms of mass density \revise{$\rho = \sum_i m_i n_i$} as 
\begin{align}
    \frac{1}{\rho} \frac{d \rho}{d r} &= - \frac{1}{T} \frac{dT}{dr} + \frac{1}{\bar m} \frac{d {\bar m}}{dr}  
    - \frac{g}{u^2_0}  - \frac{u}{u_0} \frac{du}{dr}, \label{eq:density_st} 
\end{align}
with the ideal equation of state,
\begin{align}
    P = \frac{k_{\rm B}T}{\bar{m}} \rho = u_0^2 \rho,
\end{align}
where $P$ is the pressure, $T$ is the temperature, $\bar{m}$ is the mean mass of a gas particle, $k_{\rm B}$ is the Boltzmann constant, and $u_0 (= \sqrt{k_{\rm B} T / {\bar m} })$ is the thermal velocity; 
$g$ is the gravitational acceleration, which is given by the planet's mass $M_{\rm p}$ and the gravitational constant $G$ as
\begin{equation}
    g = \frac{G M_{\rm p}}{r^2}.
\end{equation}
The temperature is calculated from the energy density and specific heat $C_{{\rm v},j}$ as
\begin{equation}
 T = \frac{e}{\sum_j n_j C_{{\rm v}, j}}. 
\end{equation}
The majority of the values of $C_{{\rm v},j}$ are taken from NASA CEA \citep[]{Gordon1994}. 
For species unavailable in NASA CEA, we adopt the specific heats for ideal gases with the internal degrees of freedom taken as three for atoms, five for diatomic molecules, and seven for all other molecules.

By use of the steady-state equation of continuity (i.e., $r^2 \rho u$ = const.), 
Equation~(\ref{eq:density_st}) can be written in terms of the bulk velocity $u$ as
\begin{align} 
    \left( 1 - \frac{u^2}{u^2_0} \right) \frac{1}{u} \frac{du}{dr} &= 
    \frac{1}{T} \frac{dT}{dr} - \frac{1}{\bar m} \frac{d {\bar m}}{dr} + \frac{g}{u^2_0} - \frac{2}{r}, \label{eq:wind}
\end{align}
which is referred to as the wind equation \citep[e.g.,][]{Parker1958}.
We integrate Equation.~(\ref{eq:wind}) inward from the upper boundary, 
which we locate at the exobase. 
We adopt the Jeans effusion velocity ${\mathcal U}$ as the outward velocity at the exobase \citep[e.g.,][]{Chamberlain1963}:
\begin{equation} \label{eq:effusion}
    \mathcal{U}_{j} = u_{0, j} \frac{(1 + X_{j}) \exp({-X_{j}})}{2 \sqrt{\pi}},
\end{equation}
where $u_{0, j}$ and $X_{j}$ are the thermal velocity and Jeans escape parameter of gas species $j$ at the exobase, respectively; $X_j$ is defined by
\begin{equation} \label{eq:e_parameter}
    X_{j} = \frac{ r_{\rm exo} g_{\rm exo} m_{j} }{k_{\rm B} T_{\rm exo}},
\end{equation}
where the subscript `exo' means the exobase.
Then, the bulk velocity is estimated by density weighted Jeans velocity:  
\begin{equation}
    u = \frac{1}{\rho} \sum_j n_j m_j \mathcal{U}_j.
\end{equation}
At the upper boundary of the exobase, the bulk velocity is replaced with the Jeans effusion velocities of the individual species in Equation.~(\ref{eq:continuous}).

\subsection{Chemical reactions} \label{sec:chemical reaction}

We consider both thermo- and photochemical reactions.  
We adopt the chemical network provided in \cite{Johnstone2018}, which is validated to reproduce the properties of the Earth's upper atmosphere and has been used in several previously published atmospheric models.
The chemical network, however, does not include some reactions that are expected to play a crucial role in high XUV environments of interest in this study.

The total heating rate due to chemical reactions, $Q^{\rm chem}$, is calculated as 
\begin{equation} \label{eq:Q_chem}
    Q^{\rm chem} = \sum_{\ell} E_{\ell} R_{\ell},
\end{equation}
where $E_\ell$ is the enthalpy of reaction for the $\ell$th reaction 
and $R_\ell$ is the reaction rate, which is calculated as
\begin{equation}
    R_\ell = k_\ell \prod_{j} n_{j};
\end{equation}
the RHS means the reaction coefficient, $k_\ell$, times the product of the number densities of all of the reactants.
We calculate the source/sink term of Equation.~(\ref{eq:continuous}) by summing all reactions, including the chemical and photochemical reactions.

\subsubsection{Thermochemistry \label{sec:thermochem}}


We consider thermochemical reactions between 47 chemical species composed of the four elements, H, C, O, and N, including ions and electrons. 
We adopt all of the reactions and species
from \citet{Johnstone2018}.
A major difference from \cite{Johnstone2018} and other previous models is that we also include chemical reactions involving internal excitation and ionization. 
The former is detailed in Section~\ref{sec:thermochem}.1, while for the latter, we take into account ionization via collision with electrons, following \cite{Voronov1997}, who presents fitting formulae for the reaction coefficients. 
In addition, unlike previous models, we consider the effects of endothermic reactions (Section~\ref{sec:thermochem}.2) and recombination and emissive de-excitation (Section~\ref{sec:thermochem}.3).

\begin{center}
    \ref{sec:thermochem}.1. \, \textit{Excitation and De-excitation}
\end{center}

We consider the excitation and de-excitation of N, O, and O$^+$ via collisions with electrons, following \cite{Tayal2006} and \cite{Ito2021}. 
Those rate coefficients are estimated with the effective collision strength, $\gamma$. 
As for excitation, the rate coefficient, $k_{lu}$, is given by \citep[e.g.,][]{Tielens2005}
\begin{equation}
    k_{lu} = \gamma \frac{8.629 \times 10^{-6}}{s_l \sqrt{T}} 
    \exp \left(- \frac{\Delta E_{lu}}{k_{\rm B}T} \right), \label{eq:excitation}
\end{equation}
where the subscripts $l$ and $u$ denote the lower and upper states, respectively, $s_l$ is the statistical weight of state $l$, and $\Delta E_{lu}$ is the energy difference between the upper and lower states.
As for de-excitation, the rate coefficient is given by 
\begin{equation}
    k_{ul} = \frac{s_{l}}{s_{u}} k_{lu} 
    \exp \left(\frac{\Delta E_{lu}}{k_{\rm B}T} \right). \label{eq:de-excitation}
\end{equation}
We ignore the small dependence of $\gamma$ on $T$ and adopt the value of $\gamma$ at $T = 1 \times 10^4$~K, which is typical for highly irradiated upper atmospheres, following \citet[]{Ito2021}. 
Also, we include transitions between states with different electronic configurations, namely, $^4$S-$^2$D-$^2$P for N, $^3$P-$^1$D-$^1$S for O, and $^3$P-$^1$D-$^1$S for O$^+$. 
The detailed level parameters of the statistical weight and energy and transition parameters are summarized in Appendix~\ref{Ap:Spectro}.

\begin{center}
    \ref{sec:thermochem}.2. \, \textit{Endothermic reactions}
\end{center}

We consider the reverse reactions for the reactions in the chemical network of \cite{Johnstone2018}. 
Unlike in the atmospheres of the present-day solar-system planets, the upper atmosphere of interest here is sufficiently hot that endothermic reactions are expected to occur efficiently. 
The rate coefficients of reverse reactions, $k_r$, are derived according to the principle of microscopic reversibility, which is discussed in several studies focusing on the atmospheric composition of hot Jupiters \citep[e.g.,][]{Visscher2011, Heng2016} and are given by  
\begin{equation}
    k_{r} = k_{f} \exp \left(- \frac{\Delta G}{k_{\rm B} T} \right) 
            \left( \frac{k_{\rm B}T}{P_0} \right)^{q_{p} - q_{r}}, 
\end{equation}
where $k_f$ is the rate coefficient of the forward reaction, $\Delta G$ is the Gibbs free energy change for the forward reaction at the standard pressure ($P_0$ = $1 \times 10^6$~dyn/cm$^2$), and $q_{p}$ and $q_{r}$ are the stoichiometric coefficients of products and reactants in the forward reaction, respectively.
We calculate $\Delta G$ for a given temperature from the enthalpy and entropy taken primarily from NASA CEA. 
For the values unavailable in NASA CEA, we adopt the values for ideal gases with the internal degrees of freedom taken as three for atoms, five for diatomic molecules and seven for other molecules. 
For their specific entropy at the standard state, we use the values calculated by the Gaussian-2 composite method, which are available in the NIST Computational Chemistry Comparison and Benchmark Database \citep[]{Johnson1999}. 
We calculate the enthalpy for excited states simply by calculating the sum of the excitation energy and ground-state enthalpy, for which no measurements are available.

\begin{center}
    \ref{sec:thermochem}.3. \, \textit{Recombination and Emissive De-excitation}
\end{center}

Our chemical network also includes radiative recombination and de-excitation via spontaneous emission. 
We assume that the radiation emitted upon recombination directly escapes to space without extinction by the atmospheric gas.

As for the spontaneous emission, we approximate the effects of radiative transfer by estimating the frequency-integrated probability for photons to escape from the atmosphere \citep[known as the escape probability method;][]{Irons1978}, instead of using a detailed treatment of radiative transfer. 
The escape probability $P_e$ is given as a function of optical thickness $\tau$ \citep[]{Hollenbach1979, Kwan1981}: 
\begin{equation}
    P_e =     
    \begin{cases}
    \sqrt{\pi} \tau
        \left(a_e + \displaystyle{\frac{\sqrt{\ln \tau}}{1+b_e\tau}} 
        \right)^{-1} & (\tau \geq 1), \\ \\
        \displaystyle{\frac{1-\exp(-2\tau)}{2\tau}} & (\tau < 1),
    \end{cases} \label{eq:eprob} 
\end{equation} 
where $a_e = 1.2$ and $b_e = 1.0 \times 10^{-5}$. 
The upper atmosphere of interest here is sufficiently tenuous that the line profile is determined by the Doppler (thermal) broadening. 
The frequency-integrated optical thickness is thus given by $\tau = \sqrt{\pi} \tau_0$, with the optical thickness at the line center $\tau_{0}$, which we estimate using the Einstein $A$ coefficients. 
In this method, it is assumed that the line profile of radiation, which is determined by the temperature of the source gas, never varies during propagation.

We assume that the radiation propagates equally upward and downward (i.e., a two-stream approximation).
The total escape probability $P_{\rm tot}$ is, thus, given by
\begin{equation} 
    P_{\rm tot} = \frac{P_e (\tau) + P_e (\tau_{\rm tot} -\tau) }{2}, \label{eq:eprob_total}
\end{equation}
where $\tau$ and $\tau_{\rm tot}$ are the optical thicknesses at the emission altitude and the lower boundary, respectively. 
We regard the photons absorbed below the lower boundary to be escaped. 
This is because those absorbed photons are quickly partitioned into the internal energy of the atmospheric gas and are, then, re-radiated by a blackbody emission, since the lower atmosphere is in the local thermal equilibrium. 
Thus, we calculate the transition rate via spontaneous emission with the Einstein $A$ coefficient multiplied by the total escape probability.

\subsubsection{Photochemistry \label{sec:photochem}}

We use the photochemical network presented in \citet{Johnstone2018}. 
We exclude the reactions  numbered ~446, 447, 454, 455, 459, and 468 in the appendix of \citet[]{Johnstone2018}, for which the absorption cross-sections or quantum yields are unavailable in their references.
For all the reactions, we take the wavelength-dependent absorption cross-sections from the PHIDRATES database \citep[]{Huebner2015}.
Our model covers the wavelength range between 0.5~nm and 400~nm to allow us to resolve the oxygen chemistry.

The reaction rate coefficient of the $\ell$th photoreaction is given by
\begin{equation}
    k_\ell = \int_{\lambda^{{\rm cr}}_\ell}^{\infty} \sigma_\ell (\lambda) I_{\lambda} (r) \, \mathrm{d}\lambda,
\end{equation}
where $\lambda$ is the wavelength, 
$\lambda^{{\rm cr}}_\ell$ is the threshold wavelength for the $\ell$th reaction, 
$\sigma_\ell (\lambda)$ is the absorption coefficient for a given wavelength, and $I_{\lambda} (r)$ is the irradience per photon, quadratic centimeter, wavelength, and second for a given wavelength at altitude $r$.
$I_{\lambda} (r)$ is calculated by the radiative transfer equation:
\begin{equation}
I_{\lambda} (r) = I_{\lambda}(\infty) \exp \left\{ - \frac{1}{\mu} \sum_{\ell} \left(\int^\infty_r  \sigma_{\ell} (\lambda) n_{\ell} \, \mathrm{d}r \right) \right\},
\end{equation}
where $I_{\lambda}(\infty)$ is the irradiance at the top of the atmosphere, $\mu$ is the cosine of the stellar zenith angle, 
and $n_{\ell}$ is the number density of the absorbed species for the $\ell$th reaction.
As in \citet[]{Johnstone2018}, we assume $\mu = \cos (66^\circ)$ to evaluate the global averaged structure of the upper atmosphere.

The absorbed photon energy is largely consumed by the chemical reactions.
For photodissociative reactions, the remaining energy is converted into the kinetic energy of molecules and atoms, which is subsequently dissipated as heat through collisional relaxation.
We assume that these conversions occur instantaneously. 
Thus, the heating rate due to photodissociation is given by
\begin{equation} \label{eq:Q_Diss}
    Q^{\rm uv} = \sum_{\ell} \int^{\lambda^{{\rm cr}}_{\ell}}_0 \left(\frac{hc}{\lambda} - \frac{hc}{\lambda^{\rm cr}_{\ell}} \right) n_{\ell} \sigma_{\ell} (\lambda) I_{\lambda} (r) \,\mathrm{d}\lambda,
\end{equation}
where $h$ is the Planck constant and $c$ is the speed of light.
The terms with $(hc/\lambda)$ and $(hc/ \lambda^{\rm cr}_{\ell})$ represent the absorbed photon energy for a given wavelength and the energy required for the $\ell$th reaction to occur, respectively.


For photoionization reactions the remaining energy is consumed in a complex manner.
The energy is first transformed into the kinetic energy of electrons.
Those electrons, which are termed photoelectrons, have a higher energy than the ambient thermal electrons.
Due to their high energy, collisions with photoelectrons result in chemical reactions of neutral species, including secondary photoionization, and heating of thermal electrons.
For such photoelectron-induced processes, we assume a local approximation such that the produced photoelectrons are consumed locally, as assumed in \citet[]{Johnstone2018}. 
The outline of the calculation method is as follows \citep[see][for the details]{Johnstone2018}:
First we obtain the initial energy distribution of photoelectrons, assuming the energy of each photoelectron as the remaining energy from each photoionization process. 
Then, the degradation of the energies of the photoelectrons is calculated by inelastic collisions between photoelectrons and neutral species, including chemical reactions and excitation, from higher energy to lower energy.
Finally, we estimate the heating rate for the photoelectrons, $Q^{\rm pe}$, using the energy distribution calculated above in the expression of energy transfer from photoelectrons to ambient thermal electrons given by \citet[]{Schunk1978}. 
Here we consider the inelastic collision of the photoelectrons with H, C, O, N, O$_2$, N$_2$, CO and CO$_2$.
The cross sections, as a function of the photoelectron energy, are taken from \citet[]{Voronov1997} for H, \citet[]{Suno2006} for C, \citet[]{Jackman1977} for O, \citet[]{Tian2008b} for N, \citet[]{Green1972, Jackman1977} for O$_2$, \citet[]{Green1965, Green1972} for N$_2$, \citet[]{Sawada1972, Jackman1977} for CO, \citet[]{Jackman1977, Bhardwaj2009} for CO$_2$. 
Excitation states that are not included in our chemical model are assumed to be quenched immediately via radiative de-excitation.

\subsection{Diffusion and Conduction} \label{sec:diffusion conduction}
\subsubsection{Diffusion}

We consider multicomponent molecular and eddy diffusion. 
The total diffusion velocity of species $j$ is expressed as the sum: 
\begin{equation}
    u_j = u^{\rm mol}_j + u^{\rm eddy}_j, 
\end{equation}
where $u^{\rm mol}_j$ and $u^{\rm eddy}_j$ are the velocities caused by molecular and eddy diffusion, respectively. 
Although a minor-component approximation is often used in upper-atmospheric modelling \citep[e.g.,][]{Banks1973}, we should use a diffusion model applicable to a wide range of conditions because the dominant species and temperature change with XUV intensity and altitude in the upper atmosphere.

For molecular diffusion, we adopt the formula derived by \citet[]{Garcia2007}, which is also used in hydrodynamic simulations for hot Jupiters \citep[]{Garcia2007b} and for hot rocky super-Earths \citep[]{Ito2021}. 
The formula is based on the momentum equations for a multicomponent gas derived by \citet{Burgers1969}, assuming no momentum transfer via diffusion, local gas neutrality (i.e., the ambipolar constraint), and no external electromagnetic forces. 
The diffusion model explicitly includes the collisions among all species and the resultant dragging effect.
Although not repeating the explicit expression of the diffusion model here, we adopt the second-order approximation of the diffusion matrix and their binary diffusion coefficients for evaluating diffusion velocity \citep[see][for the details]{Garcia2007}.
Empirical coefficients of binary diffusion are preferentially adopted. 
When no reliable data are available, however, the binary diffusion coefficients are estimated based on the hard-sphere model for neutral-neutral and neutral-electron pairs, the Coulomb interaction model for ion-ion and ion-electron pairs, and the interaction via induced polarization potential for neutral-ion pairs. 
For neutral-ion pairs, we adopt values of the polarizability of neutral species taken primarily from the CRC Handbook \citep[]{CRC2016} or the typical value of $1.0\times 10^{-24}$~cm$^{3}$ is used when no data is available.
In addition, we include the binary diffusion coefficients for charge exchange given in Table~4.4 of \citet[]{Schunk2000}.

For eddy diffusion, the diffusion speed is given by \citep[e.g.,][]{Banks1973}
\begin{equation}
    u_{j}^{\rm eddy} = - K_{\rm E} \left( \frac{1}{n_{j}} \frac{\partial n_{j}}{\partial r} + \frac{1}{\bar H} + \frac{1}{T} \frac{\partial T}{\partial r}   \right),
\end{equation}
where $K_{\rm E}$ is the eddy diffusion coefficient and $\bar H$ is the pressure scale-height.
We adopt the conventional form of the eddy diffusion coefficient,
\begin{equation} \label{eq:eddy}
    K_{\rm E} = A_{\rm E} N^{B_{\rm E}},
\end{equation}
where $N$ is the particle number density of the entire gas and $A_{\rm E}$ and $B_{\rm E}$ are empirical constants. 
We use $A_{\rm E} = 10^8$ and $B_{\rm E} = - 0.1$, which are valid for the Earth's current atmosphere \citep[][]{Johnstone2018}.

For the diffusion of electrons, we assume that the atmosphere maintains local electrical neutrality.
Assuming no magnetic field, the electron diffusive flux is equal to the sum of the ion's diffusive fluxes \citep[e.g.,][]{Shinagawa1989}:
\begin{equation}
    u_{\rm e} n_{\rm e} = \sum_j u_{j} n_{j}.
\end{equation}
This condition is also applied
to the Jeans effusion velocity of electrons.

\subsubsection{Conduction}
Heat transport occurs via thermal conduction.
The rate of heating via conduction due to both the molecular and eddy diffusion (or molecular and eddy conduction) is given by \citep[]{Banks1973, Hunten1974}
\begin{equation} 
    Q^{\rm diff} =   \frac{1}{r^2} \frac{\partial}{\partial r}\left[ r^2 \kappa_{\rm mol} \frac{\partial T}{\partial r}
    + r^2 \kappa_{\rm eddy} \left(\frac{\partial T}{\partial r} + \frac{g}{C_P} \right) \right],
\label{eq:Qdiff}
\end{equation}
where $\kappa_{\rm mol}$ and $ \kappa_{\rm eddy}$ are the thermal conductivity due to molecular and eddy diffusion, respectively, and $C_P$ is the specific heat at constant pressure, which is derived in the same manner as $C_{\rm v}$.

The molecular conduction occurs through neutral particles, ions, and electrons. 
We adopt aconduction model for each species almost identical to that used in \citet[]{Johnstone2018}, which is outlined as follows: 
For neutral particles, we consider N$_2$, O$_2$, CO$_2$, CO, O, and H and take their conductivities from \citet[]{Schunk2000}.
We use the expression of the total conductivity given by \citet[]{Banks1973} as the mixture of neutral species.
For ions and electrons, we adopt the models given in \citet[]{Banks1973}. 
We use the number density weighted average value of the ion conductivity. 
The electron conductivity includes a reduction in conductivity caused by collisions with neutral species for which we only consider the effects of N$_2$, O$_2$, O, and H.

The eddy conductivity is related to the eddy diffusion coefficient as $\kappa_{\rm eddy} = \rho C_{\rm P} K_{\rm E}$ \citep[]{Hunten1974} .
As found in Equation.~(\ref{eq:Qdiff}), the eddy conduction includes the convective turbulence term expressed as the dry adiabatic lapse rate, $g/C_P$.

\subsection{Radiative Cooling and Heating} \label{sec:cooling heating}

We consider the radiative cooling via atomic electronic transitions and molecular vibrational-rotational transitions.
In addition, we consider heating by the absorption of stellar near-IR radiation by molecular species such as CO$_2$ and H$_2$O.
For both of the atomic and molecular transitions, we assume the statistical equilibria \citep[e.g.,][]{Lopez2001}. 
The populations (or number densities) of respective excited states in the atomic and molecular species, hereafter simply termed the level populations, are determined via collisional and radiative excitation/de-excitation. 
In the statistical equilibrium, the level populations achieve a steady state such that
\begin{equation}
    \chi_{ij} n_j^{(x)} = 0
    \label{eq:SEC}
\end{equation}
where $\chi_{ij}$ is the matrix of excitation/de-excitation rate coefficients and 
$n_j^{(x)}$ is the population of excited state $j$ in species $x$. 
Combining the processes mentioned above together, we can express $\chi_{ij}$ as
\begin{equation}
    \chi_{ij} = C_{{\rm e},ij} - {C}_{{\rm d},ij} 
    + R_{{\rm e},ij} -  \left(AP_{\rm tot}\right)_{ij}, 
\end{equation}
where $C_{\mathrm{e},j}$ and $C_{\mathrm{d},j}$ 
are the matrices of collisional excitation and de-excitation coefficients, respectively, 
$R_{\mathrm{e},j}$ is that of radiative excitation rate coefficients, $A$ represents the Einstein coefficients, and ${P}_{\rm tot}$ the escape probability.
We solve Equation~\eqref{eq:SEC} with mass conservation and estimate the radiative cooling/heating rate $Q^{\rm rad}$ as
\begin{equation} \label{eq:Q_rad}
    Q^{\rm rad} = \sum_j n_j \sum_i E_{ij} \left\{ R_{{\rm e}, ij} - (A P_{\rm tot})_{ij} \right\}, 
\end{equation}
where $E_{ij}$ is the energy difference between the $i$ and $j$ states. 
Below we describe the detailed treatments and assumptions for evaluating the transitions of atomic and molecular species.

\subsubsection{Atomic electronic transitions} \label{sec:Atomic}

We consider radiative cooling via atomic electronic transitions of H, C, C$^+$, N, N$^+$, O, and O$^+$. 
We consider all energy levels below the wavenumber, $\lambda^{-1}$, of 100,000 cm$^{-1}$ or below the critical level above which permitted radiative transitions for C occur; beyond the critical level there are many fine energy levels, which have similar radiative properties to the critical level. 
For the radiative emission via transitions associated with the fine structure, we only consider oxygen atoms at the ground state, which are the dominant cooling process around the exobase of relatively low-temperature upper atmospheres like that of present-day Earth \citep[]{Bates1951}. 
For other transitions with fine structures, we adopt the averaged energy and effective collision strength multiplied by the statistical weight and the Einstein A coefficients summed over the fine structure. 
The level and transition parameters are taken from \citet[]{Ito2021} for O and O$^+$ and from the CHIANTI atomic database version~10 \citep[]{DelZanna2021} for the other species. 
The details of the energy and statistical weight of each level are summarized in Appendix~\ref{Ap:Spectro}.

For atomic line cooling, we consider the collisional excitation/de-excitation and radiative de-excitation and, for simplicity, ignore the radiative excitation; the latter, however, has a limited influence on the results below, because the incoming stellar line intensity is much weaker than the emitted line intensity when the atomic line cooling is the dominant cooling process.
We consider collision only with electrons, which are abundant in the atmospheres of interest, because there is insufficient knowledge of the effective collision strength of atomic and molecular species.
We also adopt the value of the effective collision strength at $T = 1 \times 10^4$~K. 
The effective collision strength and Einstein A coefficients used in our model are summarized in Appendix~\ref{Ap:Spectro}. 
The coefficients of collisional excitation and de-excitation are derived from Equations~\eqref{eq:excitation} and \eqref{eq:de-excitation}, respectively. 
The escape probability depends on the altitude distribution of level populations in atoms.
To calculate the escape probability self-consistently by an iterative method, we adopt the following simple method: 
The level populations and cooling rates at each altitude are calculated from the top to bottom boundaries. 
The downward escape probability is estimated by Equation~\eqref{eq:eprob} on the assumption that level populations below a given layer are in the LTE, while the upward escape probability is derived from the known altitude profile of level populations determined from the non-LTE statistical equilibrium conditions. 
Finally, we estimate the total escape probability, $P_{\rm tot}$, using Equation~\eqref{eq:eprob_total}. 
The LTE assumption for the downward escape probability is thought to be valid given high collisional frequency in the lower portion of the atmosphere.
Evaluating the errors from the escape probability and estimation method is beyond the scope of this study.

For O and N, we require special treatment for the derivation of the level populations, because the number densities of those atoms at the ground state and the first and second excitation states are calculated in the chemical model. 
Thus, we estimate the number densities of the atoms at other energy levels in the statistical equilibrium, assuming the number densities included in the chemical model are unchanged. 
This assumption is invalid from the viewpoint of mass conservation,  
but never affects the results because such excited atoms are much less abundant than the ground-state ones.
In addition, we consider the emission from O at 63 and 147~$\mu$m, which is due to the transitions associated with the fine structure in the ground state. 
We adopt the cooling rate derived by \citet[]{Bates1951} with the assumptions of LTE and $P_{\rm tot} = 1$.
Since we focus on highly-irradiated planets, such simplification never affects our conclusions.

\subsubsection{Molecular vibrational-rotational transition} \label{Mol_radiation}
Molecular radiative cooling by CO$_2$, NO, and H$_2$O is included, according to \citet[]{Johnstone2018}. 
Note that we follow \citet[]{Johnstone2018} almost exactly to focus on the influence of atomic radiative cooling on the upper-atmospheric structure in this study. 
The model is described briefly below.

We consider the 15~$\mu$m fundamental band for the CO$_2$ radiative cooling, including the collisional effects of CO$_2$ with O, O$_2$, N$_2$, and CO$_2$. 
The radiative excitation is included in the same manner as in \citet[]{Johnstone2018}: All of the energy of photons absorbed by CO$_2$ is assumed to be used for the excitation of the 15~$\mu$m band. 
Also, we adopt the escape probability method for the upward direction derived from radiative transfer calculations \citep[]{Kumer1974}, depending on the amount of CO$_2$ above the considered altitude. 
Meanwhile, all photons emitted in the downward direction are assumed to be absorbed completely.

For NO cooling, we consider emission in the vibrational band at 5.3~$\mu$m. 
We include the collisional effects only with O, although \citet[]{Johnstone2018} included radiative excitation via earthshine and assumed that radiative excitation never affects heating. 
This is because such an assumption leads to overestimating the cooling rate, in particular, at low temperatures.
Furthermore, we assume all photons escape to space. 
For H$_2$O cooling, we use the radiative cooling of rotational bands derived by \citet[]{Hollenbach1979}.

As for heating, we consider the absorption of stellar irradiation by ${\rm H_2O}$ and ${\rm CO_2}$ in the optical and IR. 
We ignore Rayleigh scattering, which is negligible in the IR. 
We perform radiative transfer calculations in the IR between 500 and 10000~cm$^{-1}$ (or 1.0 and 20~$\mu$m) with a resolution of 0.01~cm$^{-1}$. 
We calculate the absorption cross sections using the software package \textit{kspectrum} \citep[]{Eymet2016}, using the HITRAN2012 molecular spectroscopic database \citep[]{Rothman2013}. 
We derive the radiative cross-sections at 200~K and 1~Pa and use them throughout the atmosphere. 
We only consider the main isotopes, $^{12}$C$^{16}$O$_2$ and H$_2^{16}$O. 
As in \citet[]{Johnstone2018}, we consider only wavenumber bins with an absorption cross section above $10^{-22}$ cm$^2$, because weak lines are not absorbed in the tenuous upper atmospheres of interest in this study. 
For the stellar IR spectrum, we assume a blackbody spectrum of 5777~K, which is equivalent to the effective temperature of the present Sun. 
The choice of blackbody temperature never affects our results because 
such limited amounts of CO$_2$ and H$_2$O absorb limited IR radiation from the star.
The integrated stellar intensity is also assumed to be equal to that received by the present-day Earth.

\subsection{Hydrodynamic Cooling} \label{sec:advection}
Hydrodynamic cooling, which is the combined effects of work done by pressure, internal advective transport plus kinetic energy, 
and conversion to
the gravitational potential, is known to dominate the cooling process in the upper parts of escaping atmospheres of highly-irradiated planets \citep[e.g.,][]{Yelle2004, Tian2008a}.
Instead of the approximate formula used in \citet[]{Tian2008a}, who only consider adiabatic cooling, we adopt the standard energy equation of an inviscid fluid under the influence of planetary gravity \citep[e.g.,][]{Garcia2007}:

\begin{align} \label{eq:Q_Adiabat}
    Q^{\rm hy} = \frac{\partial [(e + \frac{1}{2} \rho u^2 + P) u r^2]}{r^2 \partial r} + \rho u g.
\end{align}
We neglect the stellar tide and the centrifugal force due to the orbital motion because the planets of interest are so far from their central star, and their exospheric radius is so small relative to their Hill radius, that those effects are negligible.

\subsection{Numerical Procedure} \label{sec:numerical}

We integrate the thermal and compositional structure of the upper atmosphere for a given XUV irradiation spectrum. 
The atmosphere is divided into spherical layers. 
The lowermost layer is 2~km thick and the layer thickness increases with altitude in such a way that the thickness ratio of two neighboring layers is 1.02.
Physical quantities for each layer are defined at the midpoint of the layer.
To determine the outer edge of the atmosphere (i.e., the exobase), we add layers upwards until the Knudsen number $\rm Kn$, which is the ratio of the mean free path length of gas particles to the local scale-height, approaches unity (exactly, $0.7 \leq {\rm Kn} \leq 1.2$).

We find steady-state solutions of the upper atmosphere structure, integrating from the lower boundary to the exobase.
We adopt the present-day Earth's lower atmosphere as the lower boundary condition; we use the values of the number density of O, O$_2$, and N$_2$ and the temperature at the altitude of 75~km taken from the empirical NRLMSISE-00 model \citep[]{Picone2002} on 1990 January 1.
In addition, we assume Earth-like CO$_2$ and H$_2$O mixing ratios of $4.0 \times 10^{-4}$ and $6.0 \times 10^{-6}$, respectively. 
The detailed lower boundary conditions used in our model are given in Table~\ref{Table:Lbound}.

\begin{table*}[t] 
\caption{Lower boundary condition} \label{Table:Lbound}
  \begin{tabular}{ccccccc} \hline
      Altitude~(km) & Temperature~(K) & [O]~(cm$^{-3}$) & [O$_2$]~(cm$^{-3}$) & [N$_2$]~(cm$^{-3}$) & [CO$_2$]~(cm$^{-3}$) & [H$_2$O]~(cm$^{-3}$) \\ \hline
        75 & 229.4 & $3.466 \times 10^{7}$ & $1.069 \times 10^{14}$ & $4.171 \times 10^{14}$ & $2.096 \times 10^{11}$ & $3.144 \times 10^{9}$ \\                    
    \hline
  \end{tabular}
\end{table*}

We adopt an implicit solver of $DLSODE$ with the backward differential formula \citep[]{Hindmarsh1983} for the time integration of Equations~\eqref{eq:continuous} and \eqref{eq:energy}.
The solver is suitable for stiff ordinary-differential-equation systems
such as the chemical network \citep[e.g.,][]{Grassi2014}. 
We adopt 10$^{-4}$ and 10$^{-7}$, respectively, as the values of the relative and absolute tolerances for the solver. 
From the lower boundary, 
where mean molecular weight, temperature, and wind profiles are known, we radially integrate Equation~(\ref{eq:density_st}). 
Allowing the steady-state density structure, we recalculate the number density of each species at each location for every time step, keeping the mixing ratio derived from the continuous equation.
For the initial condition, we assume that the atmosphere is perfectly mixed and the temperature and mixing ratios of all layers are the same as those at the lower boundary.
For the criterion
of convergence, we introduce the variable defined as
\begin{equation}
    \Delta \Theta = \frac{1}{\Theta}  \frac{d \Theta}{dt}, 
\end{equation}
where $\Theta$ is the physical quantity. 
When $\Delta \Theta$ of all of the physical quantities become less than 10$^{-6}$ in all of the layers, we judge the calculation to be converged. 

\subsection{UV Spectrum} \label{sec:parameters}

For the present-day solar spectrum, we adopt the spectrum model that \citet{Claire2012} developed based on the ATLAS~1 measurements obtained near the solar maximum \citep[]{Thuillier2004}. 
To investigate the response of the atmospheric structure to an increase in the amount of XUV radiation, we regard the present-day solar spectrum as the reference and simply multiply the XUV intensity at each wavelength by $N$ for which we consider values of 1 to 1000 in this study.
This approach is similar to that in \citet[]{Tian2008a}, who extrapolated the XUV spectrum to that in more active conditions from the observed solar minimum and maximum spectra. 
We note that the range of XUV wavelength in this method ($\leq 105$~nm) is different from the hydrogen ionization edge ($\leq 91$~nm), following the definition of \citet[]{Tian2008a}.
The actual shape of the XUV spectrum is not scaled by the amount of XUV radiation and FUV and NUV spectra also depend on the specific stellar activity \citep[e.g.,][]{Claire2012}.
To discuss the sensitivity to the UV spectrum, we also calculate the thermospheric structure using spectra emitted by young Sun-like stars \citep[]{Claire2012}.
We investigate the impact of the UV spectrum in Section~\ref{subsec:spectrum}.

\section{Results} \label{sec:result}


Here we investigate the response of the thermal structure of the upper atmosphere to an increase in the XUV irradiation level. 
As described in the introduction, the focus of this study is on the role of atomic line cooling.

\subsection{Importance of Atomic Line Cooling} \label{subsec:importance}

\begin{figure}[tb]
    \centering
    \includegraphics[width=1.0\linewidth]{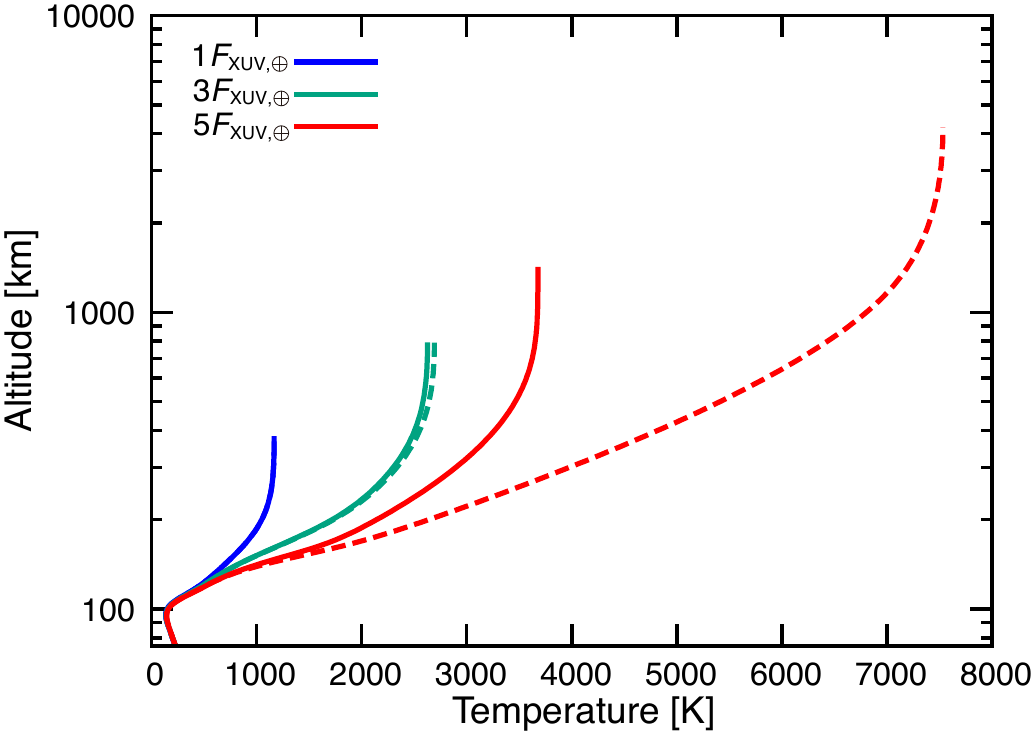}
    \caption{
        Effects of atomic line cooling on the upper-atmospheric structure. 
        Temperature profiles simulated with (\textit{solid lines}) and without (\textit{dashed lines}) atomic line cooling are shown for three different XUV irradiation levels, one times (blue), three times (green), and five times (red) the present-day Earth's one, $\Fearth$.
        }
    \label{fig:temperature_profile}
\end{figure}

Figure~\ref{fig:temperature_profile} shows the vertical temperature profiles in the upper atmosphere for three different levels of XUV irradiation ($\FXUV$), one, three, and five times the present-day Earth's irradiation($\Fearth$), which are indicated with blue, green, and red solid lines, respectively. 
For comparison, the profiles obtained without atomic line cooling are also shown by dashed lines. 
In every case, temperature is found to increase with increasing altitude because of heating at high altitudes (see below for details). 
In the cases of $1 \Fearth$ and $3 \Fearth$, the solid and dashed lines overlap each other, indicating that the atomic line cooling is ineffective. 
By contrast, in the case of $5 \Fearth$, an obvious difference is found between the solid and dashed lines; namely, the atomic line cooling results in significant temperature. 
Although not shown, the difference increases with increased XUV irradiation.

\begin{figure}[th]
    \centering
    \includegraphics[width=1.0\linewidth]{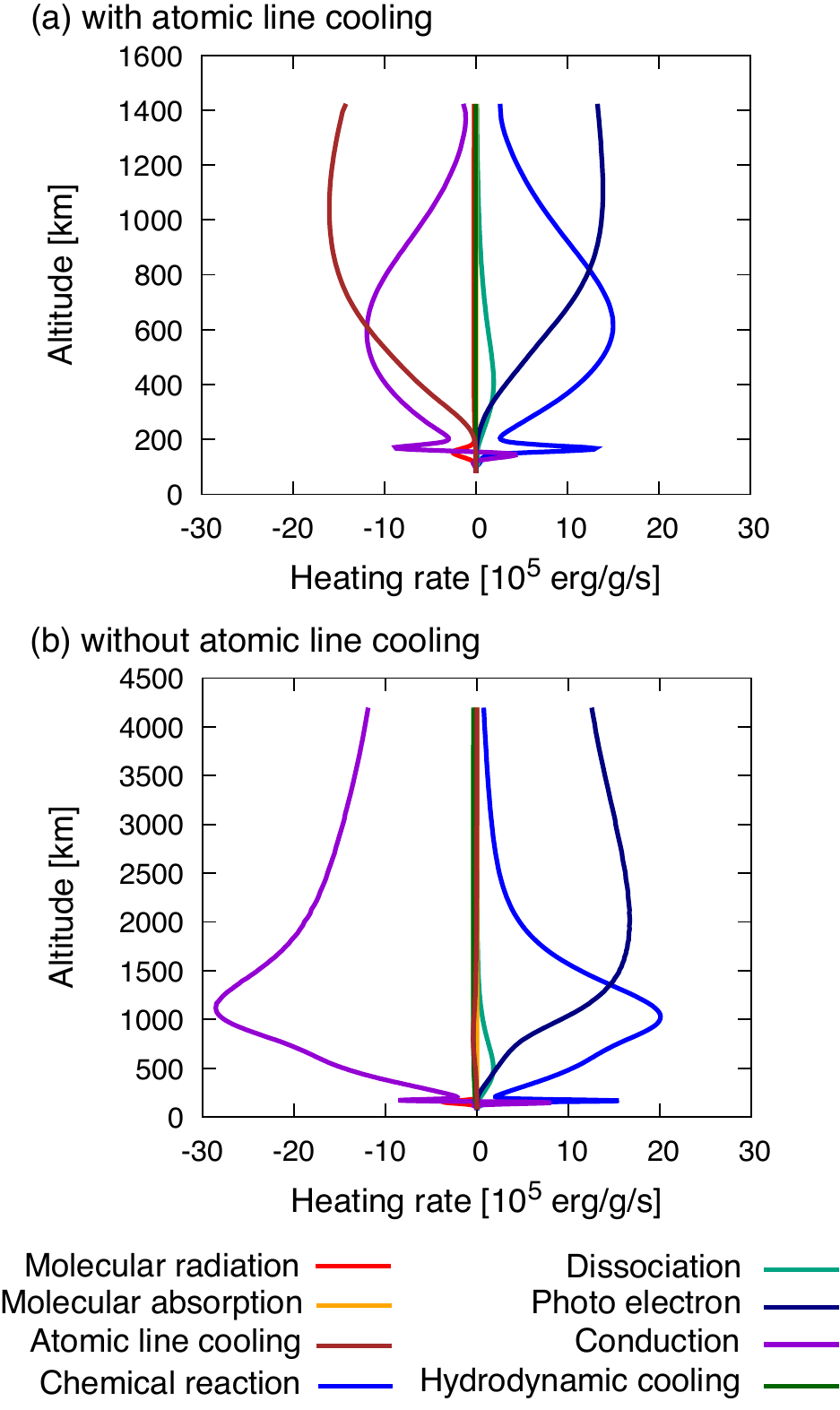}    
    \caption{
    Profiles of energy budget for the XUV irradiation five times that the present-day Earth's with (\textit{upper panel}) and without (\textit{lower panel}) atomic line cooling.
    The molecular loss (red) and absorption (yellow) of radiative energy are calculated from the first and second terms, respectively, in Equation~(\ref{eq:Q_rad}) for $Q^{\rm rad}$. 
    The atomic line cooling (brown) is also calculated from Equation~(\ref{eq:Q_rad}) for $Q^{\rm rad}$. 
    The energy budget associated with chemical reactions (blue) and photodissociation (light green) are given by $Q^{\rm chem}$ (see Equation~(\ref{eq:Q_chem})) and by $Q^{\rm uv}$ (see Eq.~(\ref{eq:Q_Diss})), respectively. 
    The rate of heating caused by photoionization (dark blue), which produces photoelectrons, is given by $Q^{\rm pe}$ (see Section~\ref{sec:photochem} for the details). 
    The energy budgets associated with conduction (purple) and adiabatic cooling (green) are given by $Q^{\rm diff}$ in Equation~(\ref{eq:Qdiff}) and $Q^{\rm hy}$ (green) in Equation~(\ref{eq:Q_Adiabat}), respectively.
    }
        \label{fig:energy_budget}
\end{figure}

To understand the effects of atomic line cooling, we show the profiles of the energy budget for $\FXUV$ = 5~$\Fearth$ in Figure~\ref{fig:energy_budget}. 
The upper and lower panels show the results obtained with and without atomic line cooling, respectively. 
In both cases, incident stellar XUV ($\lesssim$ 91~nm) is absorbed and, thereby, photoelectrons are produced at high altitudes (see the dark blue lines). 
Such photoelectrons collide with and excite the ambient atoms. 
When atomic line cooling is omitted (see Figure~\ref{fig:energy_budget}$b$), almost all of the energy thus absorbed is transferred downward by conduction (see the purple line), followed by radiative cooling via vibration of molecules such as NO and CO$_2$ at low altitudes ($\sim$ 100~km; see the red line). 
The upper atmosphere is in radiative equilibrium, as previously understood \citep[e.g.,][]{Johnstone2018}.

By contrast, in Figure~\ref{fig:energy_budget}($a$), 
the atomic line cooling (brown line) produces a major contribution to the energy loss, in addition to conduction (purple line).
Under this condition, the radiative emission from O($^1$D) associated with excitation via collisions with N$_2$, O, and e$^-$ mainly contributes to atomic line cooling. 
The dominant cooling atom and species which play a major role in collisional excitation depend on the composition and temperature of the atmospheric gas at each altitude. 
For instance, emission from excited N, N$^+$, O, and O$^+$ associated with collisional excitation by e$^-$ is dominant under highly-irradiated conditions. 
Although conduction also transfers the absorbed energy downward, similarly to the case without atomic line cooling, its contribution becomes smaller at higher altitudes, where the temperature is sufficient to excite electric transitions.
As such, the local energy loss
associated with the atomic line cooling results in a reduction in the atmospheric temperature as illustrated in Figure~\ref{fig:temperature_profile}. 
Until $\FXUV$ = 5~$\Fearth$, 
the atoms are insufficiently excited, so that the energy budget is similar to that of the case without atomic line cooling.

\subsection{Response to Increase in UV Irradiation} \label{sec:Response}
\begin{figure*}[tb]
    \centering
    \includegraphics[width=0.7\linewidth]{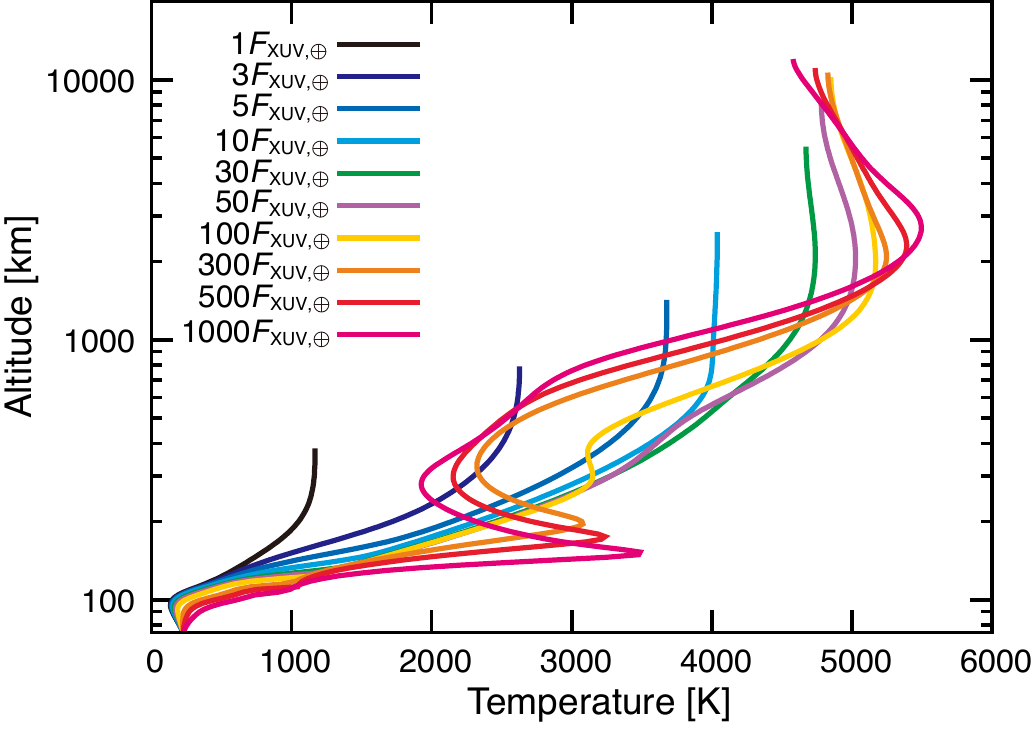}
    \caption{
    Temperature profiles for different XUV irradiation levels up to 1000 times that of the present-day Earth's, $\Fearth$. 
}
    \label{fig:temperature_profile_2}
\end{figure*}

\begin{figure*}[tb]
    \centering
    \includegraphics[width=1.0\linewidth]{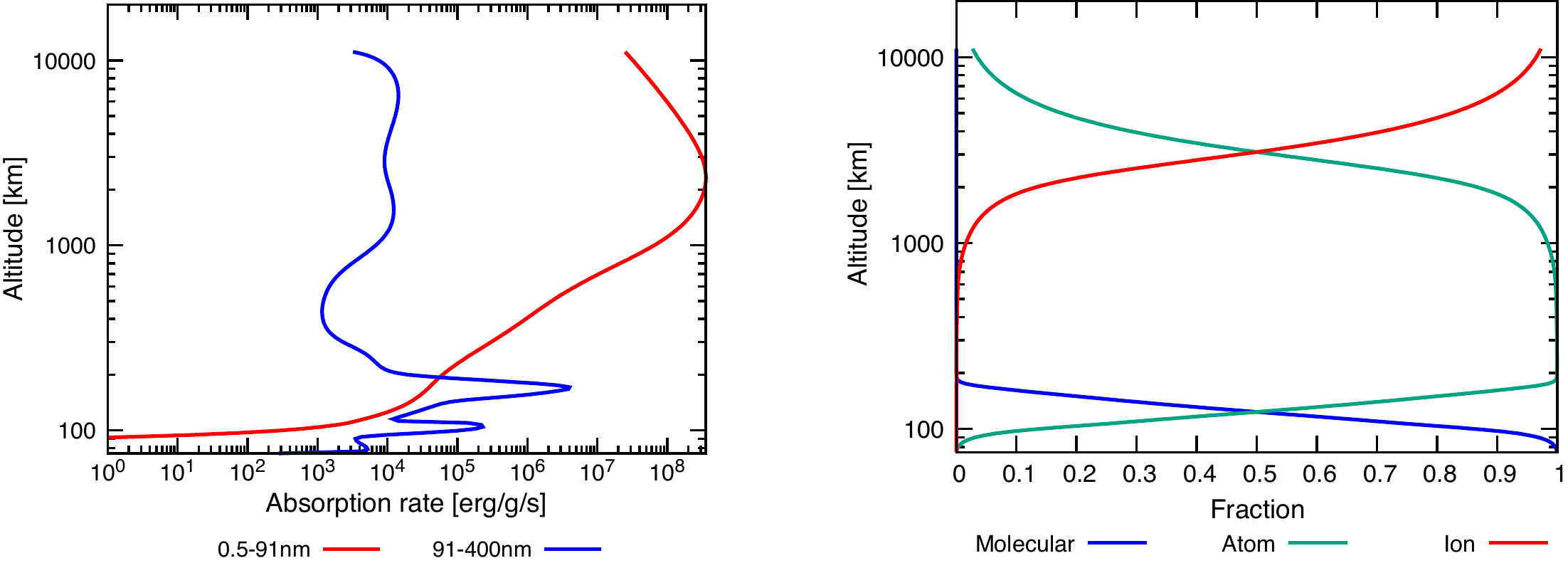}
    \caption{
    Altitude profiles of absorption and chemical properties for the XUV irradiation 500 times that of the present-day Earth's ($\FXUV$ = 500~$\Fearth$). 
    \textit{Left panel:}
    energy of radiation absorbed per unit mass per unit time in the two wavelength regions of 0.5-91~nm (red) and 91-400~nm (blue); 
    \textit{Right panel:} 
    fractions of neutral molecules (blue), neutral atoms (green), and ion species (red). 
    }
     \label{Fig:Compositional_profile2}
\end{figure*}

Figure~\ref{fig:temperature_profile_2} shows the calculated temperature profiles for the XUV irradiation level, $\FXUV$, of up to 1000~$\Fearth$. 
For $\FXUV < 300 \Fearth$ (from black to yellow lines), the temperature profiles are qualitatively similar to one another; the temperature rises monotonically with increasing altitude above $\sim$~100~km. An increase in $\FXUV$ results in an overall temperature rise, including the exospheric temperature. 
By contrast, for $\FXUV \geq 300 \Fearth$ (shown by reddish lines), the temperature profiles are not monotonic. 
Between the two stratified regions (i.e., between $\sim$~170~km and $\sim$~350~km),
the temperature falls with increasing altitude. 
Also, in the uppermost region ($\gtrsim$~2000~km), temperature falls with increasing altitude. 
In this high-$\FXUV$ regime, the exospheric temperature decreases, as the XUV irradiation increases.
This qualitative difference in the temperature profile between the low-$\FXUV$ and high-$\FXUV$ regimes can be interpreted as follows.

When $\FXUV < 300 \Fearth$, as shown in Figure~\ref{fig:energy_budget}($a$), 
the absorption of the incident stellar XUV ($\lesssim$ 91~nm), which causes photoionization, takes place at all altitudes, whereas the stellar FUV and NUV ($\sim$ 91-400~nm) are absorbed only below 100~km. 
The temperature above 100~km is thus controlled by heating via photoionization-driven
chemical reactions caused by XUV absorption. 
The specific heating rate increases with increasing altitude (see Figure~\ref{fig:energy_budget}($a$)), since the XUV energy density is lower at lower altitudes because of larger optical depths.
This is why temperature rises monotonically with increasing altitude in the low-$\FXUV$ regime.

When $\FXUV \geq 300 \Fearth$, the absorption of FUV and NUV also takes place above 100~km, in contrast to the low-$\FXUV$ regime. 
This is confirmed in Figure~\ref{Fig:Compositional_profile2}($a$), which shows the altitude profiles of the absorption rate for XUV (red) and FUV+NUV (blue). 
The contribution of FUV+NUV is much larger than that of XUV below 200~km. 
This is because neutral molecules, which absorb FUV+NUV, exist below $\sim$~200~km. This is illustrated in Figure~\ref{Fig:Compositional_profile2}($b$), where the compositional fractions of neutral molecules, neutral atoms, and ion species are shown. 
Unlike in the low-$\FXUV$ regime, because of relatively high temperatures, and thereby large scale-heights, molecules exist above 100~km in the high-$\FXUV$ regime. 
Two peaks of the absorption rate for FUV+NUV, found at $\sim$~200~km and $\sim$~100~km in Figure~\ref{Fig:Compositional_profile2}($a$), are for N$_2$ and O$_2$, respectively. 
Since the FUV+NUV energy density decreases with decreasing altitude below this, the temperature peaks at $\sim$~200~km.

Temperature also peaks at $\sim$~2000~km and then falls with increasing altitude for $\FXUV \geq 300 \Fearth$ (see Figure~\ref{fig:temperature_profile_2}). 
At 3000~km, as illustrated in Figure~\ref{Fig:Compositional_profile2}($b$), the ionized fraction is about 50.
Since ions absorb UV poorly, the absorbed energy per unit mass increases with decreasing altitude. The optical depth for XUV is approximately unity at $\sim$ 2000~km, below which it increases rapidly with decreasing altitude. 

Although thermal conduction and advective energy transport tend to relax such nonmonotonic temperature profiles, the strong dependence of atomic line cooling on temperature inhibits temperature gradients large enough for efficient energy transport.
Indeed, at $\sim$~100~km, where atomic line cooling is inefficient, while efficient absorption of long-wavelength radiation by O$_2$ occurs, eddy thermal diffusion dominates, resulting in no temperature peak. 
Thus, in systems where advection and thermal conduction work inefficiently, the atmospheric composition and the incident UV spectrum determine the profiles of the heating rate and temperature in the upper atmosphere.

\begin{figure*}[tb]
    \centering
    \includegraphics[width=0.5\linewidth]{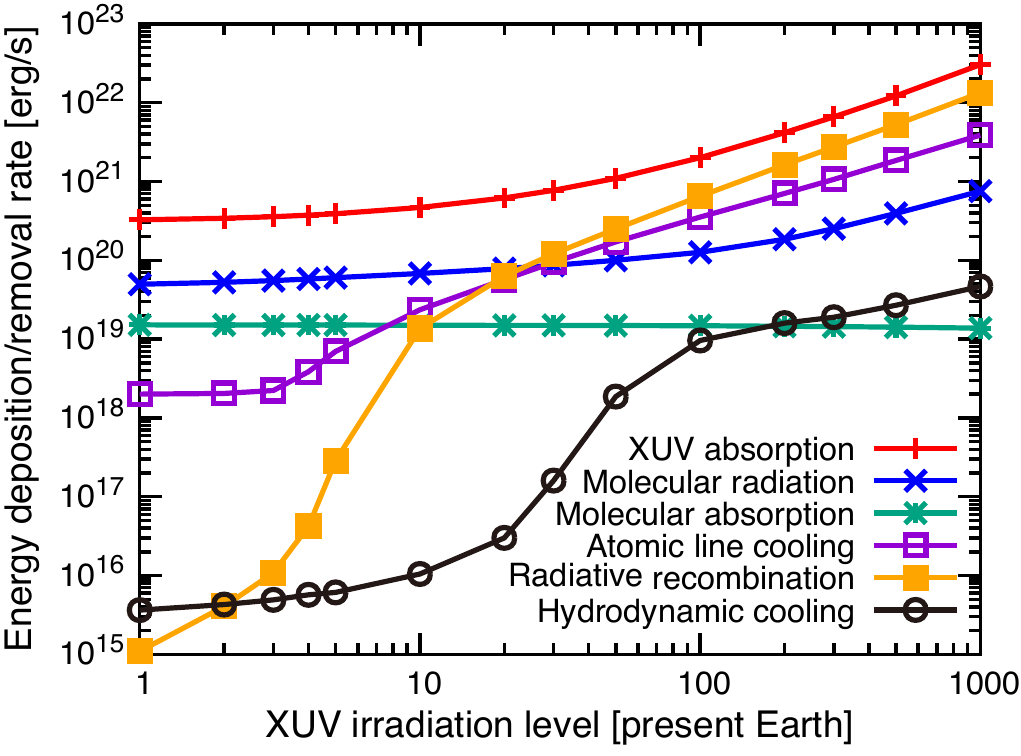}
    \caption{
     Contributions of individual energy deposition/removal rates as functions of XUV irradiation level.
     Each heating/cooling rate is integrated over the entire atmosphere.
     The red line represents the absorption in the XUV, the blue one shows the molecular cooling, the green one the molecular absorption, 
     the purple one shows the atomic line cooling, the yellow one shows the radiative recombination, and the black one shows the hydrodynamic cooling. 
     Note that the contribution of chemical reactions is not shown, because chemical reactions only change the chemical potentials and do not bring about deposition nor loss of energy from the atmosphere.
     \label{fig:energy_contribution}
     }
\end{figure*}

Figure~\ref{fig:energy_contribution} shows the rates of energy deposition/loss integrated over the entire atmosphere (i.e., $\int 4 \pi r^2 Q \mathrm{d}r$) as functions of the XUV irradiation level.
The energy deposition via XUV absorption (red line) increases with the increased XUV irradiation level; the dependence is less  than linear because X-ray ($<$~10nm) and FUV ($\gtrsim$~100~nm) are incompletely absorbed in the atmosphere. 
The energy deposition rate via molecular absorption (green line) is independent of the XUV irradiation, because the abundances of H$_2$O and CO$_2$ are also insensitive to the latter.

For energy loss, while molecular rotation-vibration cooling (blue) dominates for $\FXUV < 10 \Fearth$, atomic line cooling (purple) and radiative recombination (orange) become more effective with increasing $\FXUV$ and dominate for $\FXUV > 10 \Fearth$. 
The radiative recombination rate depends more strongly on the XUV irradiation level than on the atomic line cooling, and, consequently, the former dominates over the latter for $\FXUV \geq 30 \Fearth$. This can be explained as follows: First, the atomic line cooling rate is proportional to the number density of electrons (produced via photoionization of atoms). 
This is because under non-LTE conditions, where collisional excitation (mainly with electrons) occurs less frequently than radiative de-excitation, the former controls the atomic line cooling. Furthermore, the radiative de-excitation rate is largely insensitive to the gas species, including ionized states. 
Meanwhile, the radiative recombination rate is proportional to the square of the number density of electrons, because by definition recombination requires a collision between an electron and an ion.
Since more electrons are produced at higher XUV levels, the radiative recombination is more sensitive to $\FXUV$.
As such, the temperature rises, and thus expansion of the thermosphere, is suppressed under high-XUV conditions.

Finally we make a few comments on the above results. First, although radiative recombination dominates over atomic cooling at high-XUV irradiation levels, this does not mean that the latter is less important. It is because of the suppression of temperature rise via atomic cooling that the radiative recombination becomes effective. 
Second, for $\FXUV = 5 \Fearth$, although the atomic line cooling yields an energy-loss rate smaller by a factor of five than the molecular cooling, the former has a significant effect on the thermal structure of the upper atmosphere (see Figure~\ref{fig:temperature_profile}). 
As described in Section~\ref{subsec:importance}, this is because the local energy balance between XUV absorption and atomic line cooling hinders a rise in temperature in the upper atmosphere.
Finally, the hydrodynamic cooling always makes a limited contribution to the energy loss, as discussed in the following subsection.

\begin{figure*}[tb]
    \centering
    \includegraphics[width=0.5\linewidth]{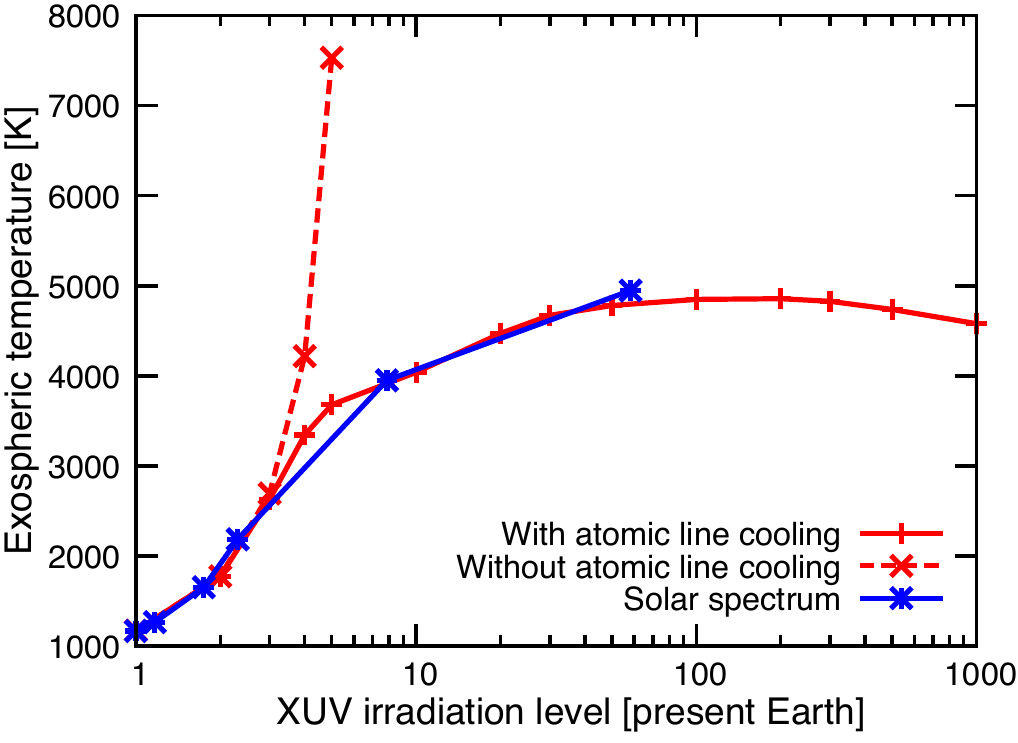}
    \caption{
        Exobase temperature as a function of XUV irradiation level, 
        $\FXUV$ (in the unit of that of the present-day Earth's, $\Fearth$). 
        The solid and dashed red lines represent the cases with and without atomic line cooling, respectively. 
        For the stellar XUV spectra, we have simply multiplied the emission intensity for the present-day Sun by $\FXUV/\Fearth$ at each wavelength. 
        The result obtained with inferred spectra for young Sun-like stars at different ages \citep[][]{Claire2012} is also shown with the blue line (see Section~\ref{subsec:spectrum}). 
    }
\label{fig:exospheric_temperature}
\end{figure*}

\subsection{Inhibition of Blow-off} \label{subsec:blowoff}

\begin{figure*}[tb]
    \centering
    \includegraphics[width=0.6\linewidth]{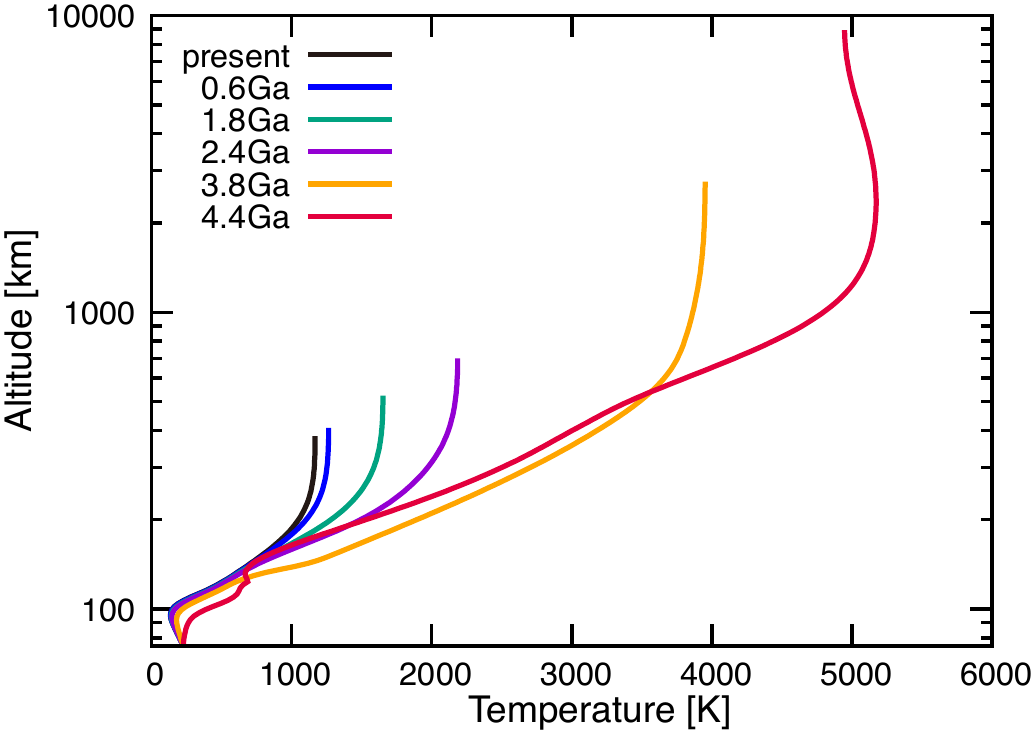}
    \caption{
     Temperature profiles for solar irradiation at different ages.
     \label{fig:Temp_differentAge}
    }
\end{figure*}

Figure~\ref{fig:exospheric_temperature} shows the exobase temperature $\Texo$ as a function of the XUV irradiation level $\FXUV$ with the effects of atomic line cooling (red solid line). 
For comparison, we also show the solutions without the effects of atomic line cooling (red dashed line). 
First, without atomic line cooling, the exobase temperature rises rapidly with the increased XUV irradiation level for $\FXUV \leq 5 \Fearth$. Beyond this level, although not shown here, an increase in XUV irradiation results in a relatively gradual decrease in exobase temperature, as found in \citet[][see their Figure~2]{Tian2008a}. 
In this regime \citep[called the hydrodynamic regime in][]{Tian2008a}, high temperature leads to large Jeans effusion velocity. 
Thus, such a decrease in exobase temperature is a hydrodynamic effect (i.e., the effect of advective cooling).
When advection is ignored (i.e., completely hydrostatic equilibrium being assumed), no solution is found for $\FXUV > 5 \Fearth$ \citep[see][]{Tian2008a}, which is often called the atmospheric blow-off.

When the atomic line cooling is incorporated (solid red line), the behavior of the exobase temperature is in clear contrast to that obtained without atomic line cooling. 
The exobase temperature increases gradually with $\FXUV$, then levels off at $\FXUV \simeq 100 \Fearth$, and, thereafter, decreases slightly with increasing $\FXUV$ for $\FXUV > 300 \Fearth$.
Of particular interest is that the atmosphere does not enter the hydrodynamic regime until $\FXUV$ = 1000~$\Fearth$. 
As $\FXUV$ increases from 1~$\Fearth$ to 1000~$\Fearth$, the number density weighted escape parameter (see Equation~(\ref{eq:e_parameter})) decreases from 106 to 8.6, indicating that advection becomes increasingly significant. 
Nevertheless, the value is still larger than the blow-off criterion \citep[$\sim$ 2.5;][]{Watson1981}. 
The decrease in exobase temperature found for $\FXUV > 300 \Fearth$ is not due to advective cooling, but due to an increase in the degree of ionization.

\subsection{Sensitivity to Stellar Spectrum} \label{subsec:spectrum}

Young Sun-like stars emit much larger amounts of XUV radiation than the present Sun and become less active with age. 
In addition to the total amount of radiation, the emission intensity at each wavelength (or emission spectrum) varies with age. 
Thus, taking the age change in stellar emission spectrum \citep[]{Claire2012} into account, we calculate and show the exobase temperature as a function of the XUV irradiation level, 
shown by the blue line in Figure~\ref{fig:exospheric_temperature}. 
As found by comparison between the blue and red solid lines, the age change in stellar spectrum has a limited effect on the $T_\mathrm{exo}$-$\FXUV$ relationship.

Figure~\ref{fig:Temp_differentAge} shows temperature profiles for solar irradiation at different ages. 
Under the same amount of XUV irradiation, the temperature profiles are similar to those obtained by linear-scaling XUV spectra shown in Figure~\ref{fig:temperature_profile_2}, as well as the exobase temperature shown in Figure~\ref{fig:exospheric_temperature}. 
Small differences are, however, found in the lower thermosphere (at $\sim 100$~km), especially for younger stellar spectra, because young Sun-like stars emit larger amounts of FUV radiation and FUV absorption mainly contributes to heating of the lower thermosphere for high-XUV conditions (see Figure~\ref{Fig:Compositional_profile2}).
As such, the detailed shape of stellar UV spectrum has a small impact on the upper thermosphere where atmospheric escape rates are determined.


%
\section{Discussion} \label{sec:discussion}
\subsection{Comparison with Previous Studies}

\begin{figure*}
    \centering
    \includegraphics[width=0.6\linewidth]{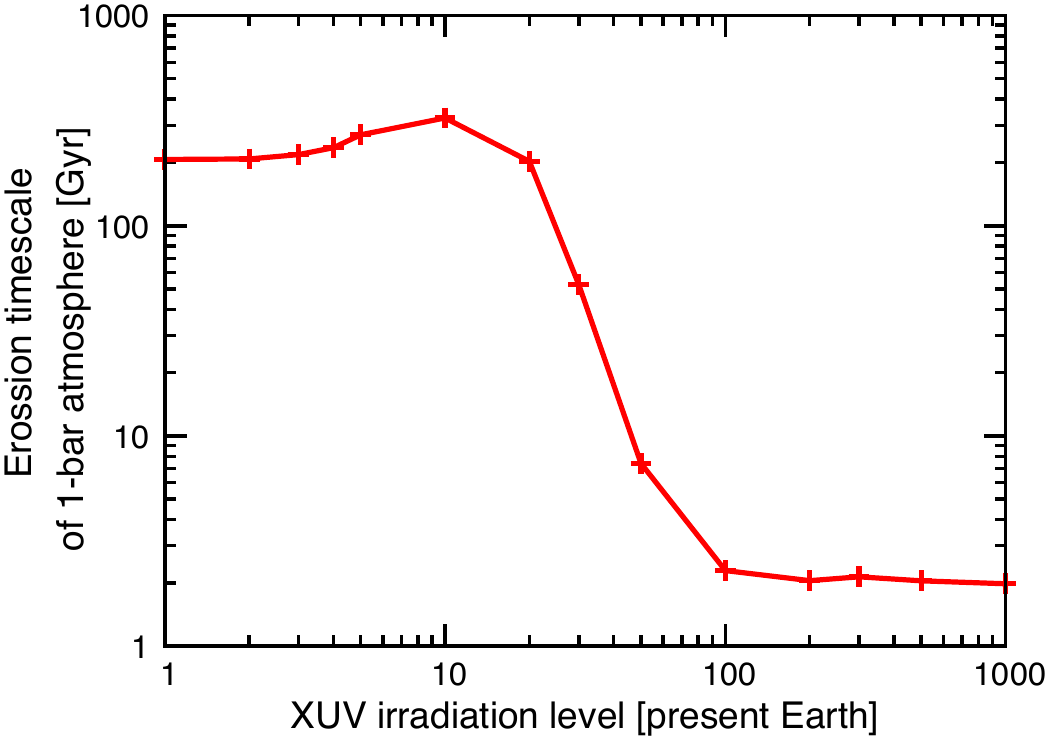}
    \caption{
    Estimated lifetime of the 1 bar N$_2$-O$_2$ Earth-like atmosphere. We have estimated the lifetime, dividing the mass of the present Earth's atmosphere by the mass-loss rate calculated from our model for each XUV irradiation flux.
    }
    \label{fig:lifetime}
\end{figure*}
The question of the vulnerability of the present-day Earth's atmosphere to strong XUV irradiation has long been explored, because the Sun is known to have emitted much stronger XUV radiation in the past than at present \citep[e.g.,][]{Ribas2005}. 
\citet{Opik1963} predicted that the Jeans escape parameter exceeds the critical value of 1.5 at a high level of XUV irradiation, leading to an uncontrolled process of atmospheric escape, and defined such a state as the atmospheric blow-off. 
With detailed treatments of photochemistry and radiative transfer and, thereby, self-consistent computations of heating efficiencies, 
\citet[]{Kulikov2007} modeled the hydrostatic equilibrium structure of the thermosphere of the terrestrial planets with enhanced XUV irradiation. 
Their model predicted a thermospheric temperature of $\sim$~10,000~K for $\FXUV \sim 10 \Fearth$, for which the escape parameter is over the blow-off threshold.

Also self-consistently computing heating efficiencies, \citet{Tian2008a} investigated the response of an Earth-like N$_2$-O$_2$ atmosphere to an increase in the XUV irradiation level. 
First they demonstrated that no hydrostatic equilibrium solution is found beyond $\FXUV \simeq 5 \Fearth$ and confirmed that the atmospheric blow-off state predicted by \citet{Opik1963} is achieved. 
Meanwhile, including the effect of advection, they did not find any uncontrolled process such as that imagined by \citet{Opik1963}. 
Beyond the \textit{hydrostatic} blow-off threshold, the adiabatic cooling due to advection (or outflow) becomes effective in keeping the thermosphere at relatively low temperatures. 
Then, as the XUV irradiation increases, advection becomes strong and the thermospheric temperature decreases. 
Consequently the atmospheric escape occurs in an energy-limited fashion \citep[][]{Tian2013}. 
Recently \citet{Johnstone2019} developed fully hydrodynamic models of planetary upper atmospheres and confirmed that the atmospheric escape occurs in the form of the transonic hydrodynamic escape (or the Parker wind) for the solar spectrum supposed at 4.4~Gyr ago.

By contrast, our results show that an N$_2$-O$_2$ Earth-like atmosphere is much more resistant to high-XUV irradiation than predicted by previous studies \citep[][]{Kulikov2007,Tian2008a,Johnstone2019}. 
This is due to the effect of atomic line cooling, as shown in Section~\ref{sec:result}. 
In our model, the thermosphere never enters such a hydrodynamic regime and remains almost in hydrostatic equilibrium until $\FXUV$ = 1000~$\Fearth$. 
This has a great impact on the loss of the atmosphere. 
For example, while \citet[]{Johnstone2019} estimated that the mass-loss rate is $1.8 \times 10^9$~g s$^{-1}$ for an N$_2$-O$_2$ Earth-like atmosphere irradiated by the 4.4~Ga Sun, our model for the same stellar spectrum estimates the mass-loss rate at $4.8 \times 10^4$~g s$^{-1}$, which is smaller by approximately four orders of magnitude than the previous value.

\subsection{Implications for the Evolution of Earth and Terrestrial Exoplanets}

A fundamental question associated with the Earth's evolution is whether the modern atmosphere would be stable in high-XUV environments of the early Earth. 
Dividing the mass of the Earth's atmosphere ($5.3 \times 10^{21}$~g) by the mass-loss rate ($1.8 \times 10^9$~g~s$^{-1}$) that \cite{Johnstone2019} obtained using \cite{Claire2012}'s solar XUV spectrum at 4.4~Ga, one can estimate the lifetime of a 1 bar N$_2$-O$_2$ Earth-like atmosphere to be 0.09~Myr. 
Such an extremely short lifetime relative to planetary evolution timescales suggests that modern Earth's atmosphere could never have survived the active phases of the Sun and must have been formed recently. 
Therefore \citet{Johnstone2019} concluded that Earth's early atmosphere was significantly different in amount and composition from the current atmosphere. 

Our model, however, predicts much longer lifetimes. 
In Figure~\ref{fig:lifetime}, we show the estimated lifetime of a 1 bar N$_2$-O$_2$ atmosphere as a function of XUV irradiation level. 
The lifetime $\tau_\text{life}$ is found to depend on the XUV irradiation level in a somewhat complicated way: for $\FXUV \lesssim 10 \Fearth$, $\tau_\text{life}$ increases slightly with increasing $\FXUV$. Then, $\tau_\text{life}$ becomes sensitive to and decreases rapidly with increasing $\FXUV$ until $\FXUV$; $100 \Fearth$. Thereafter, $\tau_\text{life}$ is almost constant for $\FXUV \gtrsim 100 \Fearth$. This dependence can be interpreted as follows, according to our simulation results. 
For low-level XUV irradiation ($\FXUV \lesssim 10 \Fearth$), only the H species such as H and H$^+$ are escaping, while the major components, O and N, are not because their escape parameters are sufficiently large. For sufficiently small values of the escape parameter, 
the H effusion velocity itself is insensitive to an increase in exobase temperature (see Equation.~(\ref{eq:effusion})), resulting in almost constant mass loss rates.
A slight decrease in mass loss rate appears because the mass loss rate is determined solely by the number density of H at the exobase, which is, by definition of the exobase, inversely proportional to the scale-height or the exospheric temperature.

For $10\Fearth \lesssim \FXUV \lesssim 100\Fearth$, by contrast, N and O are dominantly escaping, which is shown in Figure~\ref{fig:lifetime_app} of Appendix~\ref{sec:supfigure}. For such relatively high XUV irradiation, the exobase becomes comparable to the planetary radius; 
then, a rise in $\FXUV$ results in an increasing escape surface area and decreases the gravitational potential at the exobase. 
Such feedback brings about the strong dependence of the mass-loss rate on $\FXUV$.
(Note: The response of the escape velocity to the exospheric temperature and the expansion of the thermosphere is also discussed in \citet[]{Tian2008a}.)
For $\FXUV \gtrsim 100~\Fearth$, the escape rate is kept almost constant, because the atomic line cooling and radiative recombination suppress the expansion of the thermosphere, and, consequently, the effect of expansion is canceled out by the exospheric temperature due to the high ionization fraction, as discussed in Section~\ref{sec:Response}.

For the same XUV spectrum as adopted by \citet[]{Johnstone2019} ($\FXUV \approx$~60~$\Fearth$), the lifetime is estimated to be as long as 3.5~Gyr. 
Also, for $\FXUV$ = 50~$\Fearth$, the calculated mass-loss rate is $2.3 \times 10^4$~g~s$^{-1}$, corresponding to a lifetime of 7.4~Gyr, longer than the age of the Earth. 
Thus, based on the vulnerability to XUV radiation, we cannot rule out the possibility that the early Earth had the same atmosphere as today, although the early atmosphere likely differed from that of today, as suggested by other evidence \citep[e.g., the faint young Sun problem;][]{Goltblatt2009}.

To discuss the resistance of exoplanetary atmospheres to stellar XUV radiation, we would have to factor in the evolution of the latter. 
Observations of stellar X-ray emission suggest that main sequence stars become less active with age and the X-ray evolution differs from star to star. 
The difference is likely due to initial stellar rotation: 
According to \citet{Tu2015}'s calculations for solar-mass stars, 
the ``fast rotators'' younger than a few hundred megayears emit X-ray radiation hundreds of times as strong as the present Sun, while X-ray emission from the ``slow rotators'' decays rapidly and its flux decreases to tens of times the present Sun's emission in a few tens of megayears \citep[see Figure~2 of][]{Tu2015}. 
In our model for $\FXUV$ = 1000~$\Fearth$, the mass-loss rate is estimated at $8.5 \times 10^{4}$~g~s$^{-1}$, corresponding to the lifetime of 2.0~Gyr for a 1~bar N$_2$-O$_2$ Earth-like atmosphere (see Figure~\ref{fig:lifetime}). 
For such significantly long lifetimes relative to the stellar active periods, even for extremely high-XUV irradiation, our results suggest that exo-Earths orbiting in the habitable zone around Sun-like stars can retain atmospheres like the Earth's present atmosphere, regardless of stellar XUV activities.
%

As mentioned in the introduction, the initial active phases (or the saturation phases) of main-sequence stars increase with decreasing stellar mass. 
Especially for stars lighter than 0.4~$M_\odot$, the saturation phase lasts on geological timescales of gigayears and continues emitting XUV hundreds of times the solar XUV \citep[e.g., see][]{Johnstone2021}.
Thus, planets currently located in the habitable zone around M dwarfs, targeted by recent and near-future exoplanet surveys, have been exposed to strong XUV radiation. 
A 1~bar N$_2$-O$_2$ Earth-like atmospheres would be lost in the saturation phases. 
Nevertheless, N$_2$-O$_2$ atmospheres of a few bars may survive, because the saturation phase ends no later than a few gigayears, even for very-low-mass stars of $\lesssim$~0.1~$M_\odot$ \citep[see Figure~10 of][]{Johnstone2021}.


In reality, diverse atmospheric composition and pressure must be considered. 
Many pieces of geological evidence suggest that the Earth's atmosphere contained a significant amount of CO$_2$ or other greenhouse molecules in the past \citep[e.g.,][]{Catling2020}. 
Beyond the solar system, diverse water contents in terrestrial planets predicted by planet formation theories \citep[e.g., see][for recent reviews]{OBrien2018,Ikoma2018} would result in diverse amounts of atmospheric CO$_2$ for terrestrial planets in the habitable zone \citep[e.g.,][]{Nakayama2019, Krissansen-Totton2021}. 
Since molecules are capable of cooling the atmosphere through the rotation-vibration transitions, oxidizing atmospheres with larger amounts of CO$_2$ are more resistant to XUV radiation \citep[e.g.,][]{Johnstone2018}.  
As seen in Section~\ref{sec:result}, however, the molecular IR cooling makes a smaller contribution to the thermospheric structure in our model than in previous models for highly irradiated atmospheres. 
Instead, enhanced atomic line cooling from carbon atoms could lead to more efficient cooling; quantitative investigation will be explored in forthcoming studies.



\subsection{Caveats}

\subsubsection{Stability of Water Vapor Atmosphere}

Another important question regarding planetary habitability would be the stability of water vapor atmospheres under intense XUV irradiation. 
In particular, since young \revise{M dwarfs} are more luminous than the aged ones currently observed \citep[e.g.,][]{Baraffe2002}, planets currently orbiting in the habitable zone would have been exposed to more stellar radiation for a longer time than the threshold flux for the runaway or moist greenhouse \citep[][]{Ramirez2014}. 
Previous models of vapor atmospheric escape predict that such potentially habitable planets lost significant amounts of water in the past \citep[e.g.,][]{Johnstone2020}. 
Our model does not provide any conclusive predictions about the stability of water vapor atmospheres, because it includes only a small amount of H$_2$O in the atmosphere. 
Without atomic line cooling, \citet[]{Johnstone2020} predicted the thermospheric temperature to be quite high 
and no gravitational separation occurs under the XUV irradiation supposed in the saturation phase. 
Thus, the escape rate will depend on whether atomic line cooling occurs faster or slower than the gravitational separation between H and O, because atomic line cooling by lifted O would enhance the separation. 
If atomic line cooling is more efficient, vapor atmospheres avoid significant loss in such luminous phases of host stars. 
This is also a subject for future study.

\subsubsection{Nonthermal Escape}
Nonthermal escape may enhance the atmospheric loss and modify the structure of the upper atmosphere. 
\citet[]{Tian2013} predicted that under moderate XUV irradiation, nonthermal escape from the exobase level leads to shrinking of the expanded, hydrodynamic atmosphere, while keeping the total escape rate approximately constant. 
This is because the hydrodynamic process determines the overall thermospheric structure in the simulations of \citet[]{Tian2013}. 
However, our results show that instead of hydrodynamic process, the atomic line cooling and radiative recombination shape the thermosphere, even in the case with $\FXUV$ = 1000~$\Fearth$.  
Thus, when nonthermal escape dominates over thermal escape, the atmospheric escape flux would increase with XUV irradiation level until the hydrodynamic cooling becomes more effective than the atomic line cooling.
Indeed, under intense XUV irradiation, the relative importance of nonthermal escape or solar wind-induced escape \citep[e.g.,][]{Terada2009, Sakata2020} would increase. 
Such nonthermal escapes may even dominate over thermal escape, because the latter is sluggish due to atomic line cooling, as shown in previous sections. 
Detailed investigation of nonthermal escape is beyond the scope of this paper and is considered as a subject for future study.



\subsection{Implications for Exoplanet Observations}
Observational constraints on physical processes occurring in the thermosphere are crucial in further understanding of the evolution of planetary atmospheres.
The compact thermosphere resulting from atomic line cooling will be confirmed through near-future exoplanet observations. 
For instance, \citet[]{Tavrov2018} proposed a strategy for constraining the thermospheric structure for strongly XUV-irradiated terrestrial planets: 
The resonant scattering of the OI triplet at a wavelength of $\sim 130$~nm in O-rich planetary exospheres produces 
large transit depths during primary transits. 
Since the transit depth reflects the size of the thermosphere plus exosphere, future UV transit observations such as WSO-UV help us understand whether atomic line cooling is effective in actual systems.

In addition, our model predicts strong line emission in the NUV to optical wavelength range from the thermosphere of O-rich atmospheres in violent XUV environments.
For instance, assuming a globally uniform structure, we estimate that such planetary atmospheres emit 
radiation with intensity of up to $2 \times 10^{20}$~erg s$^{-1}$ at wavelengths of
630.0~nm and 636.4~nm via ${\rm O}(^1{\rm D}) \rightarrow {\rm O}(^3{\rm P})$ and 
at 557.7~nm via ${\rm O}(^1{\rm S}) \rightarrow {\rm O}(^1{\rm D})$. 
Such strong emission would be detectable during secondary transits for \revise{M dwarf} systems because \revise{M dwarfs} 
are less bright in the optical wavelength, leading to high planet-star contrast ratios. 
Quantitative investigation of the observational feasibility is beyond the scope of this paper and will be the subject of future work.  
\section{Conclusions} \label{sec:conclusion}
To answer the fundamental question of whether the present-day Earth's atmosphere could survive in the harsh XUV environments predicted for young Sun-like stars and \revise{M dwarfs}, we have developed a new model of the upper-atmospheric structure, including atomic line cooling, and investigated the stability of an N$_2$-O$_2$ Earth-like atmosphere under intense XUV irradiation, of up to 1000 times the level of present-day Earth. Our results show that the effect of atomic line cooling always dominates over the hydrodynamic effect. 
In addition, atomic line cooling is so effective in reducing the exobase temperature that the atmosphere is kept close to the hydrostatic equilibrium, and the atmospheric escape remains sluggish even under extremely strong XUV irradiation. 
This is in clear contrast to the predictions from previous studies. 
Our estimates for the Jeans escape rates of N$_2$-O$_2$ atmospheres suggest that these 1 bar atmospheres survive in the early active phases of Sun-like stars. 
Even around active late \revise{M dwarfs}, provided their atmospheric pressures are more than several bars, N$_2$-O$_2$ atmospheres could survive on
geological timescales such as the age of the Earth.
These results give new insights into the habitability of terrestrial exoplanets and the Earth's climate history.

\begin{acknowledgments}
We would like to express special thanks to the following people. 
This study was motivated by discussion with Prof.~Shingo Kameda, Dr.~Go Murakami, and Dr.~Takanori Kodama in the WSO-UV project. 
We had fruitful discussion with Dr.~Yuichi Ito especially for atomic line cooling. 
Prof.~Hitoshi Fujiwara kindly shared his numerical code with us, which greatly helped us develop our numerical code. 
Prof.~Kanako Seki and Prof.~Eiichi Tajika gave crucial scientific comments and continuous encouragement and support.
This work was supported by JSPS KAKENHI JP18H05439 and JP21H01141. 
Numerical calculations were performed in part using Oakbridge-CX at the CCS, University of Tsukuba. 

\end{acknowledgments}

%






\appendix

\section{Benchmark test} \label{Ap:benchmark}
Here, we perform a benchmark test for the validation of the upper-atmospheric model that we have newly developed in this study, by checking to see if the simulated structure of the Earth's atmosphere reproduces the observed one accurately. 
For the present Earth, we use the temperature and number density profiles of the empirical NRLMSISE-00 model \citep[]{Picone2002} observed on 1990 January 1, when the Sun was approximately at the maximum of the activity cycle.
We also adopt the UV spectrum of the present Sun at the maximum of its activity \citep[]{Claire2012} for the calculation.  

\begin{figure*}[tb]
    \centering
    \includegraphics[width=1.0\linewidth]{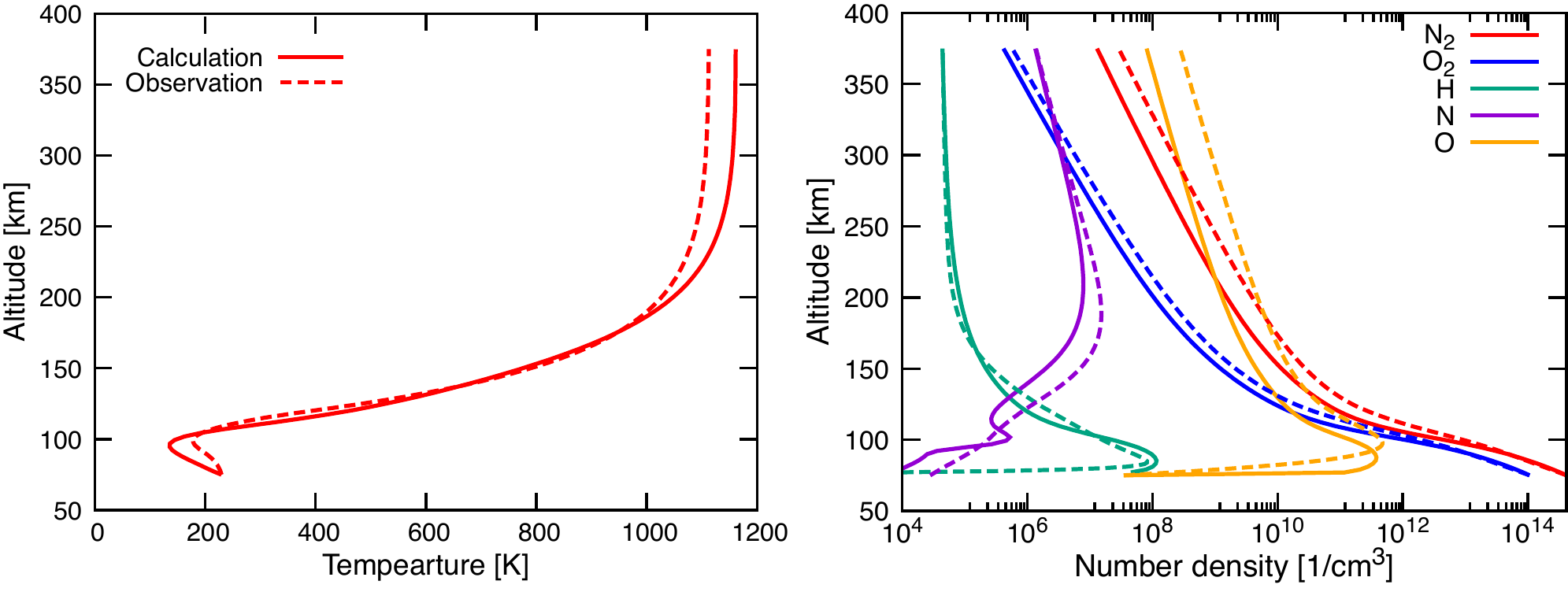}
    \caption{
    Temperature (\textit{left panel}) and compositional (\textit{right panel}) profiles for present Earth conditions.
    The solid lines are the profiles derived by our model. 
    The dashed lines represent empirical profiles derived by \citet[]{Picone2002}.
    \label{fig:Compare_Earth}
}
\end{figure*}

In Figure~\ref{fig:Compare_Earth}, we show temperature (\textit{left panel}) and compositional (\textit{right panel}) profiles for the present Earth condition. 
It can be confirmed that our model reproduces the observed profiles well. 
While the temperature profile is quite similar to the observed, some temperature differences are found in the upper and lower regions.
The difference found above $\sim$~200~km is due partly to the assumption of the common temperature between neutrals, ions, and electrons.
This assumption leads to ignoring radiative cooling by molecular vibration induced by collisions with electrons, thereby yielding higher average temperatures.
However, this assumption would be reasonable in high-XUV conditions of special interest in this study, because the temperature difference between neutrals and electrons become small relative to the average temperature \citep[]{Johnstone2018}.
The difference below $\sim$~100~km is due to the efficient eddy conduction adopted in our model; 
we use a larger value of the eddy diffusivity adopted in \citet[]{Johnstone2018} than that used in other models \citep[e.g.,][]{Roble1987}. 
Since the present Earth's atmosphere has almost hydrostatic structure, such a temperature difference in the lower atmosphere 
affects the density profile in the upper atmosphere (see Equation~\eqref{eq:density_st}). 
Thus, the calculated number densities of the main components such as O and N$_2$ differ from those observed at high altitudes. 
Nevertheless, we confirm that the calculated mixing ratio of O to N$_2$, which is more important for the thermospheric temperature, is similar to the observed one.
Thus, it would be fair to say that our model can reproduce the Earth's upper atmosphere.

\section{Supplemental figure} \label{sec:supfigure}
\begin{figure}
    \centering
    \includegraphics[width=0.6\linewidth]{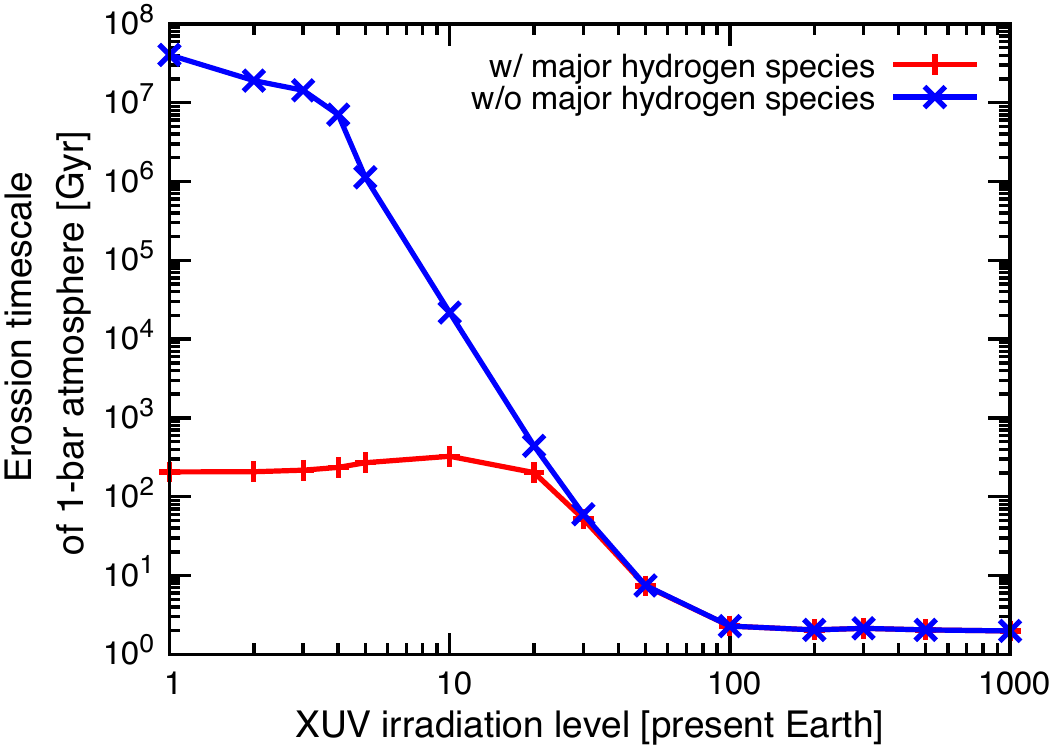}
    \caption{
    Estimated lifetime of the 1 bar N$_2$-O$_2$ Earth-like atmosphere. We have estimated the lifetime, dividing the mass of the present-day Earth's atmosphere by the mass-loss rate with (red) and without (blue) major hydrogen species (H and  H$^+$) for each XUV irradiation flux.
    The red line is exactly the same as the lifetime shown in Figure~\ref{fig:lifetime}. 
    }
    \label{fig:lifetime_app}
\end{figure}

Here, we present the supplemental figure of the lifetime of the 1 bar N$_2$-O$_2$ Earth-like atmosphere in Figure~\ref{fig:lifetime_app} to show escape species depending on XUV irradiation flux.

\section{Spectroscopic data for atomic radiative cooling} \label{Ap:Spectro}

Here, we present spectroscopic data used in our atomic line cooling model. 
Table~\ref{Table:level} shows the level structures. 
Tables~\ref{Table:transition1} to \ref{Table:transition3} show the parameters for radiative and collisional transitions.
References and derived method for level and transition parameters are given in Section~\ref{sec:Atomic}.

\begin{table}[h] 
\caption{Level structure \label{Table:level}}
\scalebox{0.9}{
  \begin{tabular}{ccccc} \hline
 Element & Level Index & Configuration  & Statistical weight & Excitation energy (cm$^{-1}$) \\ \hline
  H & 1 & 1s($^2$S) & 2 & 0.0  \\ 
    & 2 & 2s($^2$S) & 2 & 82258.96  \\   
    & 3 & 2p($^2$P) & 6 & 82259.17  \\       
    & 4 & 3s($^2$S) & 2 & 97492.22  \\           
    & 5 & 3p($^2$P) & 6 & 97492.29  \\               
    & 6 & 3d($^2$D) & 10 & 97492.34  \\                   
    \\
  C & 1 & 2p$^2$($^3$P) & 9 & 29.59122  \\       
    & 2 & 2p$^2$($^1$D) & 5 & 10192.67  \\         
    & 3 & 2p$^2$($^1$S) & 1 & 21648.04  \\             
    & 4 & 2s2p$^3$($^5$S) & 5 & 33735.22  \\                 
    & 5 & 2p3s($^3$P) & 9 & 60373.01  \\                             
    \\
  C$^+$ & 1 & 2s$^2$2p($^2$P) & 6  & 42.26666  \\         
        & 2 & 2s2p$^2$($^4$P) & 12 & 43035.75  \\           
        & 3 & 2s2p$^2$($^2$D) & 10 & 74931.60  \\                   
        & 4 & 2s2p$^2$($^2$S) & 2 & 96493.70  \\                               
        \\    
  N & 1 & 2s$^2$2p$^3$($^4$S) & 4 & 0.0  \\ 
    & 2 & 2s$^2$2p$^3$($^2$D) & 10 & 19227.95  \\     
    & 3 & 2s$^2$2p$^3$($^2$P) & 6  & 28838.51  \\    
    & 4 & 2s$^2$2p$^3$($^3$P)3s($^4$P) & 12  & 83335.60  \\        
    & 5 & 2s$^2$2p$^3$($^3$P)3s($^2$P) & 6  & 86192.79  \\            
    & 6 & 2s2p$^4$($^4$P) & 12  & 88132.45  \\                
    & 7 & 2s$^2$2p$^3$($^3$P)3s($^2$S) & 2   & 93581.55  \\                    
    & 8 & 2s$^2$2p$^3$($^3$P)3s($^4$D) & 20  & 94837.78  \\
    & 9 & 2s$^2$2p$^3$($^3$P)3s($^4$P) & 12  & 95509.79  \\
    & 10 & 2s$^2$2p$^3$($^3$P)3s($^4$S) & 4  & 96759.84  \\
    & 11 & 2s$^2$2p$^3$($^3$P)3s($^2$D) & 10 & 96833.50  \\    
    & 12 & 2s$^2$2p$^3$($^3$P)3s($^2$P) & 6  & 97793.96  \\       
    & 13 & 2s$^2$2p$^3$($^1$D)3s($^2$D) & 10 & 99663.62  \\            
    \\
  N$^+$ & 1 & 2s$^2$2p$^2$($^3$P) & 9 & 85.22956  \\     
        & 2 & 2s$^2$2p$^2$($^1$D) & 5 & 16455.11  \\       
        & 3 & 2s$^2$2p$^2$($^1$S) & 1 & 33218.58  \\               
        & 4 & 2s2p$^3$($^5$S) & 5  & 45486.12  \\                       
        & 5 & 2s2p$^3$($^3$D) & 15 & 94115.17  \\   
        \\
  O & 1 & 2s$^2$2p$^4$($^3$P) & 9  & 76.83111  \\ 
    & 2 & 2s$^2$2p$^4$($^1$D) & 5  & 15868.34  \\     
    & 3 & 2s$^2$2p$^4$($^1$S) & 1  & 33792.22  \\    
    & 4 & 2s$^2$2p$^3$($^4$S)3s($^5$S) & 5  & 73767.79  \\        
    & 5 & 2s$^2$2p$^3$($^4$S)3s($^3$S) & 3  & 76795.82  \\            
    \\
  O$^+$ & 1 & 2s$^2$2p$^4$($^3$P) & 4  & 0.00000  \\ 
    & 2 & 2s$^2$2p$^4$($^1$D) & 10  & 27826.09  \\     
    & 3 & 2s$^2$2p$^4$($^1$S) & 6  & 42125.60  \\    
    \hline
  \end{tabular}
  }
\end{table}

\begin{table}[h] 
\caption{Radiative and collisional transitional parameters \label{Table:transition1}}
\begin{tabular}{cccccc} \hline
      Element & Index of Lower Level & Index of Upper Level & Einstain Coefficient (1/s) & Effective Collision Strength \\ \hline
 H & 1 & 2 & $ 2.50 \times 10^{-6}$ & $2.42 \times 10^{-1}$  \\
   & 1 & 3 & $ 6.26 \times 10^{8}$ & $5.00 \times 10^{-1}$  \\   
   & 1 & 4 & 0.0 & $6.37 \times 10^{-2}$  \\      
   & 1 & 5 & $ 1.67 \times 10^{8}$ & $1.18 \times 10^{-2}$   \\      
   & 1 & 6 & 0.0 & $6.25 \times 10^{-2}$  \\         
   & 2 & 4 & 0.0 & $ 2.43 $  \\               
   & 2 & 5 & $ 2.24 \times 10^{7} $ & $ 4.79 $   \\         
   & 2 & 6 & 0.0 & $7.02$  \\                  
   & 3 & 4 & $ 6.31 \times 10^{6}$ & 0.0  \\            
   & 3 & 6 & $ 6.47 \times 10^{7}$ & 0.0   \\      
   \\
 C & 1 & 2  & $2.43 \times 10^{-4}$ & $1.21 $  \\                               
    & 1 & 3  & $2.13 \times 10^{-3}$ & $7.39 \times 10^{-2}$  \\                               
    & 1 & 4  & $2.15 \times 10^{1}$ & $8.03 \times 10^{-1}$  \\        
    & 1 & 5  & $3.53 \times 10^{8}$ & $6.28 \times 10^{-1}$  \\      
    & 2 & 3  & $6.38 \times 10^{-1}$ & $3.90 \times 10^{-1}$  \\
    & 2 & 4  & 0.0 & $2.09 \times 10^{-5}$  \\ 
    & 2 & 5  & $3.53 \times 10^{4}$ & $6.58$  \\  
    & 3 & 4  & 0.0 & $8.48 \times 10^{-5}$  \\ 
    & 3 & 5  & $2.54 \times 10^{3}$ & $4.58 \times 10^{-1}$  \\     
    & 4 & 5  & 0.0 & $ 9.02 \times 10^{-2}$  \\         
    \\
C$^+$ & 1 & 2 & $ 4.57 \times 10^{1}$ & $6.57$  \\
    & 1 & 3 & $ 2.90 \times 10^{7}$ & $2.92$  \\                  
    & 1 & 4 & $ 4.62 \times 10^{9}$ & $8.76 \times 10^{-1}$ \\  
    & 2 & 3 & 0.0 & $1.94 $  \\                               
    & 2 & 4 & 0.0 & $9.61 \times 10^{-1}$ \\                                     
    & 3 & 4 & 0.0 & $8.47 \times 10^{-1}$ \\   
    \\
 O & 1 & 2 & $ 8.57 \times 10^{-3}$ & $2.93 \times 10^{-1}$  \\                  
   & 1 & 3 & $ 7.87 \times 10^{-2}$ & $3.23 \times 10^{-2}$   \\                   
   & 1 & 4 & $ 1.84 \times 10^{3}$ & $2.32 \times 10^{-1} $   \\                   
   & 1 & 5 & $ 5.64 \times 10^{8}$ & $3.53 \times 10^{-1} $   \\                      
   & 2 & 3 & $ 1.26 $ & $8.83 \times 10^{-3}$   \\                   
   & 2 & 4 & $ 1.36 $ & $ 5.00 \times 10^{-1} $   \\                   
   & 2 & 5 & $ 1.75 \times 10^{3}$ & $8.23 \times 10^{-4}$  \\    
   & 3 & 4 & $ 0 $ & $ 5.00 \times 10^{-1}$   \\                   
   & 3 & 5 & $ 6.20 \times 10^{-2}$ & $ 5.00 \times 10^{-1}$  \\       
   & 4 & 5 & 0 & $ 5.00 \times 10^{-1} $  \\          
   \\
 O$^+$ & 1 & 2 & $ 7.68 \times 10^{-5}$ & 1.33   \\                 
   & 1 & 3 & $ 4.51 \times 10^{-2}$ & $4.06 \times 10^{-1}$   \\          
   & 2 & 3 & $ 9.68 \times 10^{-2}$ & 1.70   \\      
    \hline
  \end{tabular}
\end{table}

\begin{table}[h] 
\caption{Radiative and collisional transitional parameters \label{Table:transition2}}
  \begin{tabular}{ccccc} \hline
      Element & Index of Lower Level & Index of Upper Level & Einstain Coefficient (1/s) & Effective Collision Strength \\ \hline
  N & 1 & 2 & $1.30 \times 10^{-5}$ & $5.61 \times 10^{-1}$  \\                               
    & 1 & 3 & $5.22 \times 10^{-3}$ & $1.64 \times 10^{-1}$  \\                                  
    & 1 & 4 & $3.99 \times 10^{8}$ & $3.47 \times 10^{-1}$  \\  
    & 1 & 5 & $4.24 \times 10^{4}$ & $3.20 \times 10^{-2}$  \\     
    & 1 & 6 & $1.46 \times 10^{8}$ & $ 4.58 \times 10^{-1}$  \\        
    & 1 & 7 & 0.0 & $1.40 \times 10^{-2}$  \\           
    & 1 & 8 & 0.0 & $ 9.80 \times 10^{-2}$  \\              
    & 1 & 9 & 0.0 & $ 5.90 \times 10^{-2}$  \\                 
    & 1 & 10 & 0.0 & $ 1.27 \times 10^{-1}$   \\                    
    & 1 & 11 & 0.0 & $ 2.70 \times 10^{-2}$  \\                       
    & 1 & 12 & 0.0 & $ 2.00 \times 10^{-2}$  \\                          
    & 1 & 13 & $6.02 \times 10^{2}$ & $ 2.00 \times 10^{-3}$  \\                             
    & 2 & 3  & $8.47 \times 10^{-2}$ & $ 4.37 \times 10^{-1}$  \\                                
    & 2 & 4  & $9.87 \times 10^{3}$ & $ 3.35 \times 10^{-1}$  \\                                   
    & 2 & 5  & $3.35 \times 10^{8}$ & $ 2.75 \times 10^{-1}$  \\                                      
    & 2 & 6  & $4.11 \times 10^{3}$ & $ 9.78 \times 10^{-1}$  \\                                         
    & 2 & 7  & 0.0 & $ 2.20 \times 10^{-2}$  \\
    & 2 & 8  & 0.0 & $ 9.46 \times 10^{-2}$  \\
    & 2 & 9  & 0.0 & $ 4/38 \times 10^{-2}$  \\
    & 2 & 10 & 0.0 & $ 6.80 \times 10^{-3}$  \\
    & 2 & 11 & 0.0 & $ 2.16  \times 10^{-1}$  \\   
    & 2 & 12 & 0.0 & $3.28 \times 10^{-2}$  \\      
    & 2 & 13 & $3.39 \times 10^{8}$ & $2.36 \times 10^{-1}$ \\
    & 3 & 4  & $2.62 \times 10^{3}$ & $ 2.43 \times 10^{-1}$  \\                               
    & 3 & 5  & $4.83 \times 10^{7}$ & $3.12 \times 10^{-1}$  \\                                  
    & 3 & 6  & $1.08 \times 10^{3}$ & $ 6.32 \times 10^{-1}$  \\  
    & 3 & 7  & 0.0 & $ 2.03 \times 10^{-2}$  \\                                        
    & 3 & 8  & 0.0 & $ 7.50 \times 10^{-2}$  \\ 
    & 3 & 9  & 0.0 & $ 5.23 \times 10^{-2}$  \\    
    & 3 & 10 & 0.0 & $ 1.33 \times 10^{-2}$  \\       
    & 3 & 11 & 0.0 & $ 5.93 \times 10^{-2}$  \\          
    & 3 & 12 & 0.0 & $ 1.69 \times 10^{-1}$  \\             
    & 3 & 13 & $5.31 \times 10^{7}$ & $ 1.12 \times 10^{-1}$  \\      
    & 4 & 5  & 0.0 & 4.67  \\   
    & 4 & 6  & 0.0 & $ 1.31 \times 10^{1}$  \\      
    & 4 & 7  & $1.42 \times 10^{4}$ & $ 4.42 \times 10^{-1}$  \\
    & 4 & 8  & $2.54 \times 10^{7}$ & $ 1.95 \times 10^{1}$  \\   
    & 4 & 9  & $3.06 \times 10^{7}$ & $ 1.18 \times 10^{1}$  \\      
    & 4 & 10 & $3.73 \times 10^{7}$ & 2.82  \\         
    & 4 & 11 & $7.88 \times 10^{4}$ & 1.74  \\            
    & 4 & 12 & $1.43 \times 10^{3}$ & $ 9.63 \times 10^{-1}$  \\               
    & 4 & 13 & 0.0 & $ 7.48 \times 10^{-1}$  \\          
    \hline
  \end{tabular}
\end{table}

\begin{table}[h] 
\caption{Radiative and collisional transitional parameters \label{Table:transition3}}
  \begin{tabular}{ccccc} \hline
      Element & Index of Lower Level & Index of Upper Level & Einstain Coefficient (1/s) & Effective Collision Strength \\ \hline
 N & 5 & 6 & 0.0 & 1.40  \\ 
   & 5 & 7 & $9.48 \times 10^{6}$ & 6.57  \\                  
   & 5 & 8 & $1.35 \times 10^{2}$ & 3.98  \\                     
   & 5 & 9 & $3.41 \times 10^{3}$ & 2.18  \\
   & 5 & 10 & $1.30 \times 10^{5}$ & $ 7.33 \times 10^{-1}$  \\   
   & 5 & 11 & $2.56 \times 10^{7}$ & $ 1.69 \times 10^{1}$  \\      
   & 5 & 12 & $3.24 \times 10^{7}$ & 6.55  \\         
   & 5 & 13 & 0.0 & $ 6.17 \times 10^{-1}$  \\    
   & 6 & 7  & $9.28 \times 10^{-1}$ & $ 1.72 \times 10^{-1}$  \\            
   & 6 & 8  & $8.39 \times 10^{5}$ & $ 1.05 \times 10^{1}$  \\               
   & 6 & 9  & $5.87 \times 10^{5}$ & 4.17  \\                  
   & 6 & 10 & $3.85 \times 10^{6}$ & 3.31  \\                     
   & 6 & 11 & $9.07 \times 10^{3}$ & $ 8.49 \times 10^{-1}$  \\                        
   & 6 & 12 & $2.18 \times 10^{3}$ & $ 3.76 \times 10^{-1}$ \\ 
   & 6 & 13 & 0.0 & $ 1.42 \times 10^{-1}$  \\    
   & 10 & 13 & $1.44 \times 10^{1}$ & 0.0  \\ 
   & 11 & 12 & $8.69 \times 10^{3}$ & 0.0  \\       
   & 12 & 13 & $1.31 \times 10^{3}$ & 0.0  \\                
   \\
 N$^+$ & 1 & 2 & $3.90 \times 10^{-3}$ & 1.38  \\      
       & 1 & 3   & $3.20 \times 10^{-2}$ & $ 8.00 \times 10^{-1}$  \\            
       & 1 & 4   & $1.63 \times 10^{2}$ & $ 5.20 \times 10^{-1}$  \\                       
       & 1 & 5   & $3.77 \times 10^{8}$ & 1.76  \\                                  
       & 2 & 3   & $1.14 \times 10^{0}$ & 5.12  \\                                             
       & 2 & 4   & 0.0 & $8.63 \times 10^{-3}$  \\      
       & 2 & 5   & 0.0 & 1.68  \\
       & 3 & 4   & 0.0 & $3.23 \times 10^{-3}$  \\           
       & 3 & 5   & 0.0 & $1.04 \times 10^{-1}$  \\                      
       & 4 & 5   & 0.0 & $1.04 $  \\                                 
    \hline
  \end{tabular}
\end{table}

\clearpage
\bibliography{reference}{}
\bibliographystyle{aasjournal}



\end{document}